\documentclass[
11pt,
showpacs, showkeys,
amsmath,amssymb,
aps,
pre,
floatfix,
]{revtex4-1}

\pdfoutput=1

\usepackage{hyperref}
\usepackage{amsmath}
\usepackage{amsfonts}
\usepackage{amssymb}
\usepackage{bm}
\usepackage{graphicx}
\usepackage{siunitx}
\usepackage{etoolbox}
\usepackage{color}

\makeatletter
\newcommand*{\centerfloat}{%
  \parindent \z@
  \leftskip \z@ \@plus 1fil \@minus \textwidth
  \rightskip\leftskip
  \parfillskip \z@skip}
\makeatother

\apptocmd\normalsize{%
 \abovedisplayskip=10pt plus 2pt minus 8pt
 \abovedisplayshortskip=0pt plus 3pt
 \belowdisplayskip=10pt plus 2pt minus 8pt
 \belowdisplayshortskip=7pt plus 3pt minus 4pt
}{}{}

\global\long\def\V#1{\boldsymbol{#1}}

\global\long\def\D#1{\Delta#1}
\global\long\def\d#1{\delta#1}

\global\long\def\abs#1{\left|#1\right|}

\global\long\def\av#1{\left\langle #1\right\rangle }

\newcommand{\modified}[1]{#1}
\newcommand{\deleted}[1]{}
\newcommand{\Donev}[1]{}

\newcommand{\commentout}[1]{}

\newcommand{\dlapl}{{\nabla^2_\mathrm{d}}}                
\newcommand{\divg}{{\V\nabla\cdot}}                       
\newcommand{\ddivg}{{\V{\nabla}_\mathrm{d}\cdot}}         
\newcommand{\ddt}{{\frac{d}{dt}}}                         
\newcommand{\pddt}{{\frac{\partial}{\partial t}}}         
\newcommand{\paren}[1]{{(#1)}}                            
\newcommand{\Var}[1]{{\mathrm{Var}\left[#1\right]}}       
\newcommand{\Cov}[2]{{\mathrm{Cov}\left[#1,#2\right]}}    
\newcommand{\equald}{{\stackrel{\mathrm{d}}{=}}}          

\begin{document}

\title{\modified{Stochastic Simulation of Reaction-Diffusion Systems:\\ A Fluctuating-Hydrodynamics Approach}}

\author{Changho Kim$^{1}$, Andy Nonaka$^{1}$, John B. Bell$^{1}$, Alejandro L. Garcia$^{2}$ and Aleksandar Donev$^{3}$}
\affiliation{$^1$ Computational Research Division, Lawrence Berkeley National Laboratory \\
 1 Cyclotron Road, Berkeley, CA 94720, \modified{USA} \\ }
\affiliation{$^2$ Department of Physics and Astronomy, San Jose State University \\
 1 Washington Square, San Jose, CA 95192, \modified{USA} \\ }
\affiliation{$^3$ Courant Institute of Mathematical Sciences, New York University \\
 251 Mercer Street, New York, NY 10012, \modified{USA} \\ }

\begin{abstract}
We develop numerical methods for \modified{stochastic} reaction-diffusion systems based on \modified{approaches used for} fluctuating hydrodynamics (FHD).
\modified{For hydrodynamic systems, the FHD formulation is formally described by stochastic partial differential equations (SPDEs).  In the reaction-diffusion systems
we consider, our model becomes} similar to the reaction-diffusion master equation (RDME) description 
when our SPDEs are spatially discretized and reactions are modeled as a source term having Poisson fluctuations. 
However, unlike the RDME, \modified{which becomes prohibitively expensive for increasing number of molecules, our FHD-based} description naturally extends from the regime where fluctuations 
are strong, i.e., each \modified{mesoscopic} cell has few (reactive) molecules, to regimes with moderate or weak fluctuations, and ultimately to the deterministic limit.
By treating diffusion implicitly, we avoid the severe restriction on time step size that limits all methods based on explicit treatments of diffusion, and construct numerical methods 
that are more efficient than RDME methods, without compromising accuracy.
Guided by an analysis of the accuracy of the distribution of steady-state fluctuations for the linearized reaction-diffusion model, we construct several two-stage (predictor-corrector) 
schemes, where diffusion is treated using a stochastic Crank--Nicolson method, and reactions are handled by the stochastic simulation algorithm of Gillespie or a weakly second-order 
tau leaping method.
We find that an implicit midpoint tau leaping scheme attains second-order weak accuracy \modified{in the linearized setting},
and gives an accurate and stable structure factor for a time step size 
an order of magnitude larger than the hopping time scale of diffusing molecules.
We study the numerical accuracy of our methods for the Schl\"ogl reaction-diffusion model both in and out of thermodynamic equilibrium.
We demonstrate and quantify the importance of thermodynamic fluctuations to the formation of a two-dimensional Turing-like pattern, and examine the effect of fluctuations on 
three-dimensional chemical front propagation.
By comparing \modified{stochastic} simulations to deterministic reaction-diffusion simulations, we show that fluctuations accelerate pattern formation in spatially 
homogeneous systems, and lead to a qualitatively-different disordered pattern behind a traveling wave.
\end{abstract}

\date{\today}

\maketitle

\section{\label{sec_intro}Introduction}

While deterministic reaction-diffusion models have been successfully applied to explain various spatiotemporal phenomena such as pattern formation, and to gain insight into \modified{nonequilibrium transitions}, it is now widely appreciated that spatiotemporal fluctuations in the concentration of chemical species play an essential role.
Such internal or thermodynamic fluctuations, which arise from both reaction and diffusion processes, have molecular origin; microscopically, those processes occur through the movement and collision of individual molecules under thermal fluctuations.
Hence, the deterministic macroscopic description eventually fails at smaller scales where the fluctuations are significant, and a stochastic mesoscopic description is needed.
Examples include fluctuation-induced instabilities~\cite{KesslerLevine1998}, reversal of direction of front propagation~\cite{KhainLinSander2011}, violation of the law of mass action~\cite{WinklerFrey2012}, long-time tails in kinetics~\cite{GopichOvchinnikovSzabo2001}, emergence of new steady states~\cite{TogashiKaneko2004} and patterns~\cite{WangFuXuOuyang2007}, acceleration of pattern formation~\cite{LemarchandNowakowski2011}, \modified{enhanced induction time for ignition~\cite{PeetersBarasNicolis1990}, and the onset of homogeneous oscillations~\cite{MalekMansourDethierBaras2001}}.
Due to a small number of proteins involved in cellular functions~\cite{FedoroffFontana2002}, processes in cell biology are good examples~\cite{DoubrovinskiHoward2005,FangeElf2006,ZonMorelliTanaseNicolaWolde2006,FleggRudigerErban2013} where the stochastic reaction-diffusion description provides an indispensable modeling tool~\cite{BurrageBurrageLeierEtAl2011,MahmutovicFangeBergElf2012}.


A microscopic picture of reaction-diffusion, dating back to Smoluchowski~\cite{Smoluchowski1917}, assumes that molecules undergo independent Brownian motions and reactions can occur only when two molecules are close to each other.
Based on this picture, the particle-based approach to simulate a reaction-diffusion system tracks the trajectories of diffusing molecules and uses the intermolecular distance to determine whether a reaction occurs.
Exact sampling of the Smoluchowski model can be performed by first-passage kinetic Monte Carlo type algorithms~\cite{DonevBulatovOppelstrupGilmerSadighKalos2010,TakahashiTanaseNicolaWolde2010,MauroSigurdssonShrakeAtzbergerIsaacson2014}; approximate reactive Brownian dynamics (BD) using a fixed time step size forms another class of algorithms~\cite{AndrewsBray2004,LipkovaZygalakisChapmanErban2011}.
While molecular schemes, such as molecular dynamics (MD) and direct simulation Monte Carlo (DSMC), can be used for reaction-diffusion problems~\cite{DziekanLemarchandNowakowski2012,DziekanHansenNowakowski2014}, they are computationally even more expensive.
Hybrid methods combining particle and coarse-grained descriptions, either using operator splitting~\cite{ChoiMauryaTartakovskySubramaniam2012} or domain decomposition~\cite{FleggChapmanErban2012,HellanderHellanderLotstedt2012}, have also been proposed.


For the mesoscopic description of a reactive system, the master equation approach is commonly used. 
For a well-mixed (i.e., spatially homogeneous) system, the time evolution of the system (i.e., the number of molecules of each chemical species) is described by the chemical master equation (CME).
Exact sampling of the CME can be performed by the stochastic simulation algorithm (SSA) of Gillespie~\cite{Gillespie1976}, whereas the tau leaping method~\cite{Gillespie2001} can be employed as an approximate algorithm with a given time step size.
Several variants of these methods have been proposed~\cite{GillespieHellanderPetzold2013,SzekelyBurrage2014}.
For a spatially inhomogeneous system, the time evolution of the system is commonly described by the reaction-diffusion master equation (RDME), which is also known as the multivariate master equation~\cite{Gardiner2004,ErbanChapmanMaini2007}.
In this approach, the system is divided into homogeneous subsystems or cells and the number of molecules of each chemical species in each cell is tracked.
Changes in the molecule numbers occur either through hopping events of a molecule between adjacent cells or though chemical reactions within a cell.
Hopping events correspond to diffusive transport and are treated as first-order reactions.
Since the RDME is a spatial extension of the CME, exact sampling of the RDME can be performed by SSA-type algorithms~\cite{MalekMansourHouard1979,ElfEhrenberg2004}, which are called inhomogeneous SSA (ISSA).

While conceptually simple and still widely used~\modified{\cite{WangFuXuOuyang2007,LemarchandNowakowski2011,LuisHitaOrtizdeZarate2013}}, the traditional approach of solving the RDME by ISSA has the computational issue that the method becomes prohibitively slow as the number of molecules or cells increases.
Since the cell volume should be chosen sufficiently small to ensure homogeneity over each cell, large, finely-resolved grids are required for two- or three-dimensional problems.
As the spatial resolution increases, the time interval between successive events becomes very short due to rapid diffusive transfer, and hopping events greatly outnumber reaction events, which slows down ISSA~\cite{GillespieHellanderPetzold2013}. 
Several approaches have been proposed to improve the performance of stochastic sampling of the RDME, such as the next subvolume method~\cite{ElfEhrenberg2004} and its parallel simulation version~\cite{WangHouXingYao2011}.
Various implementations of the tau leaping method in a spatial context~\cite{MarquezLagoBurrage2007,RossinelliBayatiKoumoutsakos2008,IyengarHarrisClancy2010}, and the time-dependent propensity for diffusion method~\cite{FuWuLiPetzold2014}, have also been proposed.
A more aggressive approach to reduce the computational cost by avoiding the sampling of the individual diffusion events is to split diffusion and reaction in each time step and to treat diffusion in a more efficient manner. 
Various sampling methods for diffusion have been proposed, including the Gillespie multi-particle method~\cite{RodriguezKaandorpDobrzynskiBlom2006}, the multinomial simulation algorithm~\cite{LampoudiGillespiePetzold2009}, the adaptive hybrid method on unstructured meshes~\cite{EngblomFermHellanderLotstedt2009,FermHellanderLotstedt2010}, and the diffusive finite-state projection algorithm~\cite{DrawertLawsonPetzoldKhammash2010,HellanderLawsonDrawertPetzold2014}.


\modified{In this paper, we propose a numerical algorithm for stochastic reaction-diffusion systems based on approaches used for 
fluctuating hydrodynamics (FHD).}
To incorporate the effects of thermal fluctuations in a fluid, in FHD one assumes that the dynamics of the fluid can be described by the usual hydrodynamic equations 
(e.g., the Navier--Stokes equations), augmenting each dissipative flux with a stochastic flux~\cite{Ottinger2005}.
Those stochastic fluxes are modeled by spatiotemporal Gaussian white noise (GWN) and the resulting governing equations are written as stochastic partial differential equations (SPDEs).
FHD was originally developed for equilibrium fluctuations by Landau and Lifshitz~\cite{LandauLifshitz1959} and its validity has been justified for nonequilibrium 
systems~\cite{Espanol1998} through the theory of coarse graining~\cite{EspanolAneroZuniga2009}.
\modified{For further discussion of FHD compared to MD, see Ref.~\cite{VoulgarakisChu2009}.}
Various extensions and generalizations of FHD theory have been developed and successfully applied to fluctuation-induced phenomena;
\modified{see Ref.~\cite{OrtizdeZarateSengers2006} and references therein.}
Recent work by the authors has focused on FHD models of hydrodynamic transport~\cite{DonevBellFuenteGarcia2011a,DonevBellFuenteGarcia2011b} in binary fluid mixtures~\cite{DonevNonakaSunFaiGarciaBell2014,NonakaSunBellDonev2015}, multiphase flows~\cite{ChaudhriBellGarciaDonev2014}, multispecies fluid mixtures~\cite{BalakrishnanGarciaDonevBell2014,DonevNonakaBhattacharjeeGarciaBell2015}, multispecies reactive mixtures~\cite{BhattacharjeeBalakrishnanGarciaBellDonev2015}, and electrolytes~\cite{PeraudNonakaChaudhriBellDonevGarcia2016}.

Compared to our previous work~\cite{BhattacharjeeBalakrishnanGarciaBellDonev2015}, where the coupling effects of fluid hydrodynamic and chemical fluctuations have been investigated, 
here we focus on reaction and diffusion and neglect all other hydrodynamic processes (advection, viscous dissipation, thermal conduction, and cross term effects).
Rather than using a Langevin description (i.e., based on Gaussian fluctuations) of chemistry, which is only valid in the limit of vanishing fluctuations~\cite{BhattacharjeeBalakrishnanGarciaBellDonev2015}, here we employ a more accurate description of reactions based on Poisson fluctuations.
As pointed out in Ref.~\cite{EspanolAneroZuniga2009}, even though a formal SPDE description is employed, the actual interpretation of FHD always requires the notion of a \emph{coarse-graining} over a certain length scale.
The FHD equations are discretized using a finite volume approach~\cite{DonevVandenEijndenGarciaBell2010,DelongGriffithVandenEijndenDonev2013} that represents the solution in terms of the average over cells, which provides an effective coarse-graining.
Therefore, reactions can be treated in a similar manner to the RDME approach when the SPDEs are spatially discretized, and integrated in time using SSA or a weakly second-order tau leaping method~\cite{HuLiMin2011,AndersonKoyama2012}.
\modified{Recent relevant work by others includes Ref.~\cite{Atzberger2010}, in which the FHD approach has been applied to reaction-diffusion systems.
However, only fluctuations arising from diffusion have been considered (i.e., no fluctuations from chemical reactions) and modeled as additive noise.
The FHD approach has been also applied to concentration fluctuations in a ternary liquid mixture in equilibrium~\cite{OrtizdeZarateLuisHitaSengers2013} and the Model H equations for binary mixtures~\cite{ThampiPagonabarragaAdhikari2011}.}

The key difference between the FHD and RDME descriptions lies in the more efficient treatment of fast diffusion.
A number of approximate numerical methods for the RDME~\cite{RodriguezKaandorpDobrzynskiBlom2006,LampoudiGillespiePetzold2009,EngblomFermHellanderLotstedt2009,FermHellanderLotstedt2010,DrawertLawsonPetzoldKhammash2010,HellanderLawsonDrawertPetzold2014} are based on operator splitting using first-order Lie or second-order Strang splitting~\cite{Strang1968}.
In Appendix~\ref{appendix_RDME_based_schemes} we review and discuss in more detail a split scheme that uses multinomial diffusion sampling~\cite{BalterTartakovsky2011} for diffusion and SSA for reactions.
These RDME-based schemes use a time step size $\D{t}$ comparable to the hopping time scale $\tau_\mathrm{h}=\D{x}^2/(2dD)$ with $d$ being the spatial dimension, $\D{x}$ being the grid spacing, and $D$ being a typical diffusion coefficient.
Even though $\tau_\mathrm{h}$ is much larger than the mean duration between successive events in ISSA, using $\D{t}$ comparable to $\tau_\mathrm{h}$ is still very restrictive for large $D$ or small $\D{x}$.
In our FHD formulation, we treat diffusion implicitly using backward Euler or Crank--Nicolson, so that the time step size can be significantly larger \modified{(e.g., an order of magnitude larger for a given accuracy tolerance)} than the hopping time scale.
\modified{Since the time steps used in RDME simulations are already (usually an order of magnitude or more) larger than those in BD simulations, our approach allows even larger time step size compared to particle-based methods.}

While the development of numerical schemes for stochastic reaction-diffusion systems described by spatiotemporal GWN dates back to the 1990s~\cite{LemarchandLesneMareschal1995,KarzaziLemarchandMareschal1996}, much of the prior work has not been guided by numerical analysis or extensive experience from deterministic computational fluid dynamics (CFD).
With the help of well-established techniques for numerical solution of PDEs and SPDEs, we construct numerical schemes in a systematic manner to ensure accuracy is maintained for a large time step size.
To this end, we employ two-stage (i.e., predictor-corrector) Runge--Kutta temporal integrators~\cite{DelongGriffithVandenEijndenDonev2013,DelongSunGriffithVandenEijndenDonev2014}.
Rather than using operator splitting, we treat reaction and diffusion together in each stage in a manner that is second-order weakly accurate for general linearized FHD equations. 
The construction of these schemes is guided by a stochastic accuracy analysis of the (static) structure factor for linearized FHD~\cite{DonevVandenEijndenGarciaBell2010,DelongGriffithVandenEijndenDonev2013}.
\modified{The structure factor is the steady-state spectrum of the concentration fluctuations, i.e., the covariance matrix in Fourier space, see Eq.~\eqref{struct_factor}.
We apply the techniques in~\cite{DonevVandenEijndenGarciaBell2010,DelongGriffithVandenEijndenDonev2013} to predict the discrete structure factors for our scheme,
and compare them to analytical predictions of the continuum structure factors for our model in the linearized setting.}

The FHD approach inherently outperforms the RDME approach as the number of molecules per cell increases in exactly the 
same way that multinomial diffusion outperforms diffusion by hopping, or tau leaping outperforms SSA.
In fact, the computational cost of FHD methods does not significantly change as the magnitude of the fluctuations changes.
This is an obvious advantage of the FHD approach since the macroscopic limit cannot be efficiently simulated by the RDME 
approach.
However, the validity of the FHD approach cannot be taken for granted when there are only a small number of molecules in 
each cell, since in FHD the number of molecules in each cell is a continuous real-valued variable, rather than a discrete 
nonnegative integer variable as in the RDME.
We investigate this issue carefully and propose techniques to improve the accuracy of the FHD description for the case of 
a small number of molecules per cell, making our numerical schemes robust even for large fluctuations.
In particular, we develop a spatial discretization that significantly mitigates nonnegativity of the species number 
densities and closely reproduces the Poisson thermodynamic equilibrium distribution for the number of molecules in a cell.
\modified{For numerical examples considered in this paper, we show that the mean number of molecules in a cell can be as 
low as 10.}
\modified{However, if one is specifically interested in systems with only a small number of molecules per cell, one should 
use an integer-based description like RDME.
Moreover, in very dilute cases, a particle-based description like BD is actually fastest since most cells will have 
essentially no molecules in them.
However, for practical stochastic simulation of reaction-diffusion systems, where the populations of chemical species may 
have different orders of magnitude, this kind of robustness is required; even if there are a large number of molecules in 
a cell, some species may have a small number of molecules.}

The rest of the paper is organized as follows.
Section~\ref{sec_background} presents the background for our approach, including the FHD description of reaction-diffusion systems and linearized analysis in a Gaussian approximation. 
Section~\ref{sec_spatial_discretization} explains how the FHD reaction-diffusion equations can be spatially discretized using a finite-volume approach.
Section~\ref{sec_temporal_integrators} presents temporal integrators for the spatially discretized equations that handle diffusion using existing FHD techniques, and treat reactions using SSA or second-order tau leaping.
Section~\ref{sec_numerics} presents simulation results of several reaction-diffusion systems.
In Section~\ref{subsec_Schlogl_res}, for testing and validation of our numerical schemes, we use a one-species Schl\"ogl model~\cite{Schlogl1972,VellelaQian2009}.
In Section~\ref{subsec_BPM}, to compare our methods to each other and to RDME-based methods, we study two-dimensional Turing-like pattern formation in the three-species Baras--Pearson--Mansour (BPM) model~\cite{BarasPearsonMalekMansour1990,BarasMalekMansourPearson1996}.
In Section~\ref{subsec_Lemarchand_AB}, to demonstrate the ability of our approach to scale to larger systems, we present numerical simulation results for three-dimensional front propagation in a two-species model~\cite{LemarchandNowakowski2011}.
In Section~\ref{sec_conclusions}, we offer some concluding remarks and suggest future research directions.
\section{\label{sec_background}Background}

In Section~\ref{subsec_FHD_description}, we present the continuous-time continuous-space FHD description of reaction-diffusion systems.
Here, we assume that fluctuations in chemistry are described by GWN (i.e., Langevin type).
A more accurate description of chemistry based on Poisson fluctuations is incorporated in the continuous-time discrete-space description in Section~\ref{sec_spatial_discretization}. 
In Sections~\ref{subsec_struct_factor} and \ref{subsec_Schlogl_bg}, we introduce the structure factor and the Schl\"ogl reaction-diffusion model, respectively.
As one of the criteria for the development and analysis of numerical schemes, later in the paper we investigate how accurately a numerical scheme produces the structure factor for the Schl\"ogl model.

In this section, we introduce several GWN vector and scalar random fields and denote them by $\V{\mathcal{Z}}(\V{x},t)=(\mathcal{Z}_1(\V{x},t),\dots,\mathcal{Z}_d(\V{x},t))$ and $\mathcal{Z}(\V{x},t)$, respectively, with additional superscripts to distinguish the different fields.  
We assume that any two distinct processes are independent and that the noise intensity of each process is normalized, $\av{\mathcal{Z}_j(\V{x},t)\mathcal{Z}_{j'}(\V{x}',t')}=\delta_{jj'}\delta(\V{x}-\V{x}')\delta(t-t')$ and $\av{\mathcal{Z}(\V{x},t)\mathcal{Z}(\V{x}',t')}=\delta(\V{x}-\V{x}')\delta(t-t')$.
Similarly, we denote GWN vector and scalar random processes by $\V{\mathcal{W}}(t)$ and $\mathcal{W}(t)$, respectively, and assume $\av{\mathcal{W}_j(t)\mathcal{W}_{j'}(t')}=\delta_{jj'}\delta(t-t')$ and $\av{\mathcal{W}(t)\mathcal{W}(t')}=\delta(t-t')$.

\subsection{\label{subsec_FHD_description}FHD Description}

We consider a reaction-diffusion system having $N_\mathrm{s}$ species undergoing $N_\mathrm{r}$ reactions in $d$-dimensional space.
By denoting the number density of species $s$ by $n_s(\V{x},t)$, the equations of FHD for $\V{n}(\V{x},t)=(n_1(\V{x},t),\dots,n_{N_\mathrm{s}}(\V{x},t))$ are written formally as the SPDEs~\cite{BhattacharjeeBalakrishnanGarciaBellDonev2015}
\begin{equation}
\label{RD_SPDE_CLE}
\pddt n_s
=\divg\left(D_s\V\nabla n_s+\sqrt{2D_s n_s}\V{\mathcal{Z}}_s^\paren{\mathrm{D}}\right)
+\sum_{r=1}^{N_\mathrm{r}} \nu_{sr}\left(a_r(\V{n})+\sqrt{a_r(\V{n})}\mathcal{Z}_r^\paren{\mathrm{R}}\right),
\end{equation}
where $D_s$ is the diffusion coefficient of species $s$, $a_r(\V{n})$ is the propensity function indicating the rate of reaction $r$, and $\nu_{sr}$ is the stoichiometric coefficient of species $s$ in reaction $r$.
In the macroscopic limit of vanishing fluctuations, Eq.~\eqref{RD_SPDE_CLE} approaches the deterministic reaction-diffusion PDE (law of large numbers), 
\begin{equation}
\label{RD_PDE}
\pddt n_s=D_s\nabla^2 n_s+\sum_{r=1}^{N_\mathrm{r}} \nu_{sr}a_r(\V{n}).
\end{equation}
We explain below how the diffusion and reaction parts are obtained by considering the diffusion-only (i.e., no-reaction) and reaction-only (i.e., well-mixed) cases.

\subsubsection{Diffusion}

The diffusion-only SPDE
\begin{equation}
\label{SPDE_diff_only}
\pddt n_s=\divg\left(D_s\V\nabla n_s+\sqrt{2D_s n_s}\V{\mathcal{Z}}_s^\paren{\mathrm{D}}\right)
\end{equation}
can be justified by considering a microscopic system where each molecule $i$ undergoes independent Brownian motion,
\begin{equation}
\label{BD_particle}
\dot{\V{x}}_{s,i}=\sqrt{2D_s}\V{\mathcal{W}}_{s,i}.
\end{equation}
In this \emph{formal} derivation~\cite{Dean1996}, one defines the instantaneous number density field $n_s(\V{x},t)=\sum_i \delta(\V{x}-\V{x}_{s,i}(t))$ and uses Ito's rule to obtain Eq.~\eqref{SPDE_diff_only}.
This equation can also be obtained from the diffusion portion of the general multispecies FHD equations~\cite{DonevNonakaBhattacharjeeGarciaBell2015,BhattacharjeeBalakrishnanGarciaBellDonev2015}, given by nonequilibrium statistical mechanics~\cite{Kuiken1994}, by assuming a dilute solution and considering only solute species.
In addition, the linearized version of Eq.~\eqref{SPDE_diff_only} can be obtained from the multivariate master equation model (i.e., diffusion by hopping) near the macroscopic limit~\cite{Gardiner2004}.
However, although relations to those equations reaffirm Eq.~\eqref{SPDE_diff_only} near the macroscopic limit, it is important to note that Eq.~\eqref{SPDE_diff_only} is formally exact even in the case where fluctuations are large, since it is simply a rewriting of Eq.~\eqref{BD_particle}, in a representation in which the particle numbering (identity) is lost~\cite{DonevVandenEijnden2014}.

We note that the FHD equations~\eqref{RD_SPDE_CLE} and \eqref{SPDE_diff_only} are not mathematically well-defined because the solution needs to be interpreted as a distribution (or a generalized function), and the square root of a distribution is not well-defined in general.
The linearized FHD equation does not suffer from such an issue and is well-defined; the problems arise due to the multiplicative noise in Eq.~\eqref{SPDE_diff_only}.
However, even though Eq.~\eqref{SPDE_diff_only} is ill-defined, it is formally consistent with the law of large numbers (given by the deterministic reaction-diffusion equation~\eqref{RD_PDE}), the central limit theorem (given by the linearized FHD equations~\eqref{linearized_SPDE}), and large deviation theory for a collection of Brownian walkers.
In this sense, Eq.~\eqref{SPDE_diff_only} is a meaningful \emph{representation} of the physical model that is useful in constructing \emph{well-defined} mesoscopic descriptions via spatial discretization of the formal SPDEs.
Compared to obtaining a mesoscopic model by directly coarse-graining the microscopic model, the spatial discretization of the SPDE is easier in general, and can be done in a systematic manner~\cite{TorreEspanolDonev2015}.

\subsubsection{Reaction}

To see how the reaction part of Eq.~\eqref{RD_SPDE_CLE} is obtained, consider a well-mixed system with volume $\D{V}$.
By assuming that the time evolution of $\V{n}(t)$ follows the CME, we express the change over the infinitesimal time interval $dt$ as follows~\cite{GillespieHellanderPetzold2013}:
\begin{equation}
\label{dns}
d n_s = n_s(t+dt)-n_s(t) = \frac{1}{\D{V}}\sum_{r=1}^{N_\mathrm{r}} \nu_{sr} \mathcal{P}(a_r(\V{n})\D{V} dt),
\end{equation}
where $\mathcal{P}(m)$ denotes a Poisson random variable having mean $m$.
Note that Eq.~\eqref{dns} is equivalent to the CME if interpreted in the Ito sense.
The specific form of the chemical rate function $a_r(\V{n})$ that we use in this work is described in Section~\ref{subsec_treatments}.
Henceforth, we will formally write Eq.~\eqref{dns} in the differential form,
\begin{equation}
\label{dnsdt}
\ddt n_s = \sum_{r=1}^{N_\mathrm{r}} \frac{\nu_{sr} \mathcal{P}(a_r(\V{n})\D{V} dt)}{\D{V} dt}.
\end{equation}
For a more mathematically precise representation, see Refs.~\cite{Plyasunov2005,Li2007}.

The chemical Langevin equation (CLE)~\cite{Gillespie2000} is obtained under the assumption that the mean number of reaction occurrences is large~\cite{GillespieHellanderPetzold2013}.
That is, the assumption enables one to replace $\mathcal{P}(m)$ by a Gaussian random variable having the same mean and variance, to give the CLE
\begin{equation}
\label{CLE}
\ddt n_s=\sum_{r=1}^{N_\mathrm{r}} \nu_{sr}\left[a_r(\V{n})+\sqrt{\frac{a_r(\V{n})}{\D{V}}}\mathcal{W}_r\right].
\end{equation}
Since reaction is assumed to be local, the reaction part of Eq.~\eqref{RD_SPDE_CLE} is obtained from spatial extension of Eq.~\eqref{CLE}.

One of the important conclusions of our previous work~\cite{BhattacharjeeBalakrishnanGarciaBellDonev2015} was that the Langevin description \eqref{CLE} is not consistent with equilibrium statistical mechanics.
Alternative formulations based on a Langevin diffusion description~\cite{OttingerGrmela1997,HanggiGrabertTalknerThomas1984} that are consistent at thermodynamic equilibrium fail to correctly model relaxation toward equilibrium~\cite{BhattacharjeeBalakrishnanGarciaBellDonev2015}.
Instead, in order to correctly capture both small fluctuations and large deviations in equilibrium and non-equilibrium contexts, one must retain a description of chemical reactions as a Markov jump process.
That is, one must describe reactions using a stochastic differential equation driven by Poisson rather than Gaussian noise.

\subsection{\label{subsec_struct_factor}Structure Factor}

The structure factor \modified{is the steady-state spectrum of the concentration fluctuations,

\begin{equation}
\label{struct_factor}
S_{s}(\V{k})=V\av{\d{\hat{n}_{s,\V{k}}} \d{\hat{n}_{s,\V{k}}}^*},
\end{equation}
i.e.,} the variance of the Fourier mode of the number density of species $s$,
\begin{equation}
\hat{n}_{s,\V{k}}(t)=\frac{1}{V}\int n_s(\V{x},t)e^{-i\V{k}\cdot\V{x}} d\V{x}.
\end{equation}
Here we have assumed a periodic domain of volume $V$, and defined $\d{\hat{n}_{s,\V{k}}}=\hat{n}_{s,\V{k}}-\av{\hat{n}_{s,\V{k}}}$, where the brackets $\av{\;}$ denote the equilibrium average.
Here we derive an analytic expression of the structure factor from the linearized FHD equation.
We assume that there is only one species, $N_\mathrm{s}=1$, and drop the subscript $s$ for species, to write Eq.~\eqref{RD_SPDE_CLE} as
\begin{equation}
\label{SPDE_one_spec_CLE}
\pddt n
=D\nabla^2n+\divg\left(\sqrt{2Dn}\V{\mathcal{Z}}^\paren{\mathrm{D}}\right)
+a(n)+\sqrt{2\Gamma(n)}\mathcal{Z}^\paren{\mathrm{R}},
\end{equation}
where
\begin{equation}
a(n)=\sum_{r=1}^{N_\mathrm{r}} \nu_r a_r(n),\quad
\Gamma(n)=\frac{1}{2}\sum_{r=1}^{N_\mathrm{r}} \nu_r^2 a_r(n),
\end{equation}
and we have expressed fluctuations arising from all reactions by a single GWN field $\mathcal{Z}^\paren{\mathrm{R}}$.
At a spatially uniform stable steady state, $n(\V{x},t)$ fluctuates around mean number density $\bar{n}\equiv\av{n}$, where
$a(\bar{n})=0$ and $a'(\bar{n})<0$.
The linearization of Eq.~\eqref{SPDE_one_spec_CLE} around this equilibrium state is given by the central limit theorem,
\begin{equation}
\label{linearized_SPDE}
\pddt n=D\nabla^2n+\sqrt{2D\bar{n}}\divg\V{\mathcal{Z}}^\paren{\mathrm{D}}
-r(n-\bar{n})+\sqrt{2\bar{\Gamma}}\mathcal{Z}^\paren{\mathrm{R}},
\end{equation}
where $r=-a'(\bar{n})>0$ is the effective reaction rate near equilibrium and $\bar{\Gamma}=\Gamma(\bar{n})$.

The Fourier transform of Eq.~\eqref{linearized_SPDE} gives
\begin{equation}
\label{linearized_FT}
\ddt\d{\hat{n}_{\V{k}}}=-D k^2\d{\hat{n}_{\V{k}}}+\sqrt{2D\bar{n}}\:i\V{k}\cdot\hat{\V{\mathcal{Z}}}_{\V{k}}^\paren{\mathrm{D}}-r\d{\hat{n}_{\V{k}}}+\sqrt{2\bar{\Gamma}}\hat{\mathcal{Z}}_{\V{k}}^\paren{\mathrm{R}}.
\end{equation}
Since Eq.~\eqref{linearized_FT} has the form of the Ornstein--Uhlenbeck equation~\cite{Gardiner2004}, the structure factor is easily obtained as
\begin{equation}
\label{Sk_exact}
S(\V{k})=\frac{D\bar{n} k^2+\bar{\Gamma}}{D k^2+r}=\frac{\bar{n} k^2+\bar{\Gamma}/D}{k^2+\ell^{-2}},
\end{equation}
where $\ell=\sqrt{D/r}$ denotes the penetration depth.
From Eq.~\eqref{Sk_exact}, we observe that there are two limiting cases.
In the small wave number limit $k\ell\ll 1$, $S(\V{k})$ becomes $\bar{\Gamma}/r$ and does not depend on diffusion.
In fact, the result $S(\V 0)=\bar{\Gamma}/r$ is also obtained from the CME assuming the whole system is well-mixed.
On the other hand, in the large wave number limit $k\ell\gg 1$, $S(\V{k})$ becomes $\bar{n}$, which is the result for the diffusion-only system.
Hence, fluctuations are reaction-dominated at a length scale larger than $\ell$ and are diffusion-dominated at a length scale smaller than $\ell$.

We also observe that if the system is in detailed balance at its steady state, i.e., it is in thermodynamic equilibrium, then $\bar{\Gamma}=\bar{n} r$ and $S(\V{k})=\bar{n}$, consistent with a product Poisson distribution with mean number density $\bar{n}$.
Therefore, in true thermodynamic equilibrium the statistics of the fluctuations are independent of any kinetic parameters, as they must be according to equilibrium statistical mechanics~\cite{Keizer1987}.
In particular, the presence of the reactions does not change the Poisson statistics of the state of thermodynamic equilibrium.
In Section~\ref{subsec_Schlogl_res}, we use this property to judge the quality of numerical schemes.

\subsection{\label{subsec_Schlogl_bg}Schl\"ogl Model}

The Schl\"ogl model~\cite{Schlogl1972,VellelaQian2009} is given by the chemical reactions for species $\mathrm{X}$,
\begin{equation}
\label{Schlogl}
2\mathrm{X} \underset{k_2}{\stackrel{k_1}{\rightleftharpoons}} 3\mathrm{X}, \quad
\varnothing \underset{k_4}{\stackrel{k_3}{\rightleftharpoons}} \mathrm{X}.
\end{equation}
Hence, we have $N_\mathrm{s}=1$, $N_\mathrm{r}=4$, $\nu_1=\nu_3=1$, $\nu_2=\nu_4=-1$, $a(n)= k_1 n^2 - k_2 n^3 + k_3 - k_4 n$, and $\Gamma(n)= \frac12\left(k_1 n^2 + k_2 n^3 + k_3 + k_4 n\right)$.
Due to the cubic nonlinearity of $a(n)$, the well-mixed system exhibits several kinds of distributions depending on the values of the rate constants.
If detailed balance is satisfied, that is, $k_1 n_\mathrm{eq}^2=k_2 n_\mathrm{eq}^3$ and $k_3 = k_4 n_\mathrm{eq}$, the system is in thermodynamic equilibrium and the distribution follows Poisson statistics with mean number density $n_\mathrm{eq}$.
Otherwise, depending on the number of real roots of $a(n)=0$, the system exhibits a monostable distribution (for a single positive root) or a bistable distribution (for three positive roots)~\cite{VellelaQian2009}. 

The structure factor of the spatially extended Schl\"ogl model can be calculated from Eq.~\eqref{Sk_exact}.
As expected from the fact that the equilibrium distribution of the system follows Poisson statistics, $S(\V{k})=n_\mathrm{eq}$ in the case of thermodynamic equilibrium.
Note, however, that having a monostable distribution does \emph{not} imply thermodynamic equilibrium.
The structure factor of the out-of-equilibrium monostable case is not flat but exhibits a transition near $k\ell\sim 1$.
For the bistable case exhibiting metastability, the linearized theory is still applicable if one looks at fluctuations around one of the two peaks.
However, in this work we focus on the equilibrium and out-of-equilibrium cases where $a(n)$ has a single positive root.
\section{\label{sec_spatial_discretization}Spatial Discretization}

In this section, we discuss spatial discretization of the FHD equation using a finite-volume approach~\cite{DonevVandenEijndenGarciaBell2010,DelongGriffithVandenEijndenDonev2013} that converts the SPDE into stochastic ordinary differential equations (SODEs) for the cell number density $n_{s,\V{i}}(t)$. 
We develop numerical schemes to solve these SODEs in Section~\ref{sec_temporal_integrators}.
In Section~\ref{subsec_spatial_discretization_diff_only}, we first discretize the diffusion-only SPDE~\eqref{SPDE_diff_only}. 
In Section~\ref{subsec_spatial_discretization_react_diff}, we add reactions and present the continuous-time discrete-space description of the reaction-diffusion system.
In Section~\ref{subsec_treatments}, we discuss techniques to handle a small number of molecules per cell.  

For simplicity, in this paper we only consider periodic systems. However, our methods can be straightforwardly generalized to standard types of physical boundary conditions (Dirichlet, Neumann or Robin).
In particular, since chemistry is local and does not require boundary conditions, one can rely on methods we have developed in prior work without chemistry; see, for example, the discussion in Ref.~\cite{UsabiagaBellDelgadoBuscalioni2012}.

\subsection{\label{subsec_spatial_discretization_diff_only}Diffusion-Only Case}

Due to the lack of regularity of $\V{\mathcal{Z}}_s^\paren{\mathrm{D}}(\V{x},t)$ in Eq.~\eqref{SPDE_diff_only}, pointwise values of $n_s(\V{x},t)$ are not physically meaningful.
Hence, we consider instead the spatial average of $n_s(\V{x},t)$ over a cell.
We partition the system domain $L_1\times L_2\times\cdots L_d$ into cells of volume $\D{V}=\D{x}_1\cdots\D{x}_d$ and denote the cell number density of species $s$ in cell $\V{i}=(i_1,\dots,i_d)$ as
\begin{equation}
n_{s,\V{i}}(t)=\frac{1}{\D{V}}\int_{\mathrm{cell}\;\V{i}}n_s(\V{x},t)d\V{x}.
\end{equation}
We denote the face of a cell using the index $\V{f}$.
If two contiguous cells have indices $\V{i}$ and $\V{i}+\V{e}_j$ (with $\V{e}_j$ being the unit vector along the $j$-axis), the face $\V{f}$ shared by the cells is denoted by $\V{i}+\frac12\V{e}_j$.

\begin{figure}
\centering
\includegraphics[width=2in]{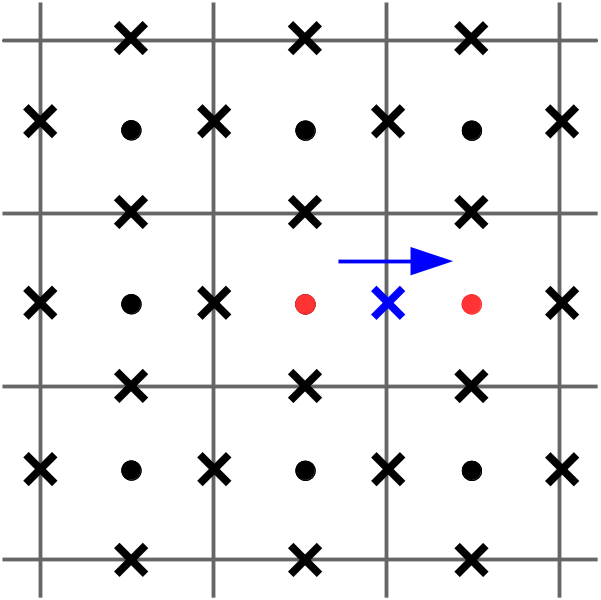}
\caption{\label{fig_cell}
Finite-volume spatial discretization in two dimensions.
The cell-averaged number density $n_{s,\V{i}}(t)$ is associated with the circles, and the face-averaged stochastic diffusive flux is associated with the crosses. The stochastic diffusive flux between the two cells having the red circles at center is depicted by the blue arrow.}
\end{figure}

To obtain a spatial discretization of Eq.~\eqref{SPDE_diff_only} that ensures discrete fluctuation-dissipation balance~\cite{DonevVandenEijndenGarciaBell2010,DelongGriffithVandenEijndenDonev2013}, we use the standard second-order discrete Laplacian operator for the deterministic diffusion part $D_s\nabla^2 n_s$ and introduce a staggered grid for the stochastic diffusive flux term $\divg\big(\sqrt{2D_s n_s}\V{\mathcal{Z}}_s^\paren{\mathrm{D}}\big)$, see Fig.~\ref{fig_cell}.
For $d=1$, a formally second-order spatial discretization of Eq.~\eqref{SPDE_diff_only} is written as 
\begin{equation}
\label{SODE_diff_only_1d}
\ddt n_{s,i}= D_s\frac{n_{s,i+1}-2n_{s,i}+n_{s,i-1}}{\D{x}^2}
+\sqrt{\frac{2D_s}{\D{V}}}\frac{\sqrt{\tilde{n}_{s,i+\frac12}}\mathcal{W}_{s,i+\frac12}-\sqrt{\tilde{n}_{s,i-\frac12}}\mathcal{W}_{s,i-\frac12}}{\D{x}}.
\end{equation}
The spatial average of $n_s(\V{x},t)$ over the interval of length $\D{x}$ around face $i\pm\frac12$ is approximated by $\tilde{n}_{s,i\pm\frac12}(t)$, whereas that of $\mathcal{Z}_s(\V{x},t)$ is modeled by $\frac{1}{\sqrt{\D{V}}}\mathcal{W}_{s,i\pm\frac12}(t)$.
To close the equation, $\tilde{n}_{s,i\pm\frac12}(t)$ is approximated by an average of $n_{s,i}(t)$ and $n_{s,i\pm 1}(t)$, that is, $\tilde{n}_{s,i\pm\frac12} = \tilde{n}\left(n_{s,i},n_{s,i\pm 1}\right)$.
Natural candidates for the averaging function $\tilde{n}(n_1,n_2)$ would be the Pythagorean means: the arithmetic, geometric, and harmonic means.
We choose a modified arithmetic average for $\tilde{n}(n_1,n_2)$ described in Section~\ref{subsec_treatments}, for reasons detailed in Appendix~\ref{appendix_averaging_function}.

Generalization of the spatial discretization~\eqref{SODE_diff_only_1d} to higher dimensions is straightforward.
For each face, a GWN process $\mathcal{W}_{s,\V{f}}$ is defined and $\tilde{n}_{s,\V{f}}(t)$ is calculated from the cell number densities of the two cells sharing the face by using the averaging function $\tilde{n}(n_1,n_2)$, see Fig.~\ref{fig_cell}.
By introducing notations $n_s(t)\equiv\{n_{s,\V{i}}(t)\}$, $\mathcal{W}_s(t)\equiv\{\mathcal{W}_{s,\V{f}}(t)\}$, and $\tilde{n}_s(t)\equiv\tilde{n}\left(n_s(t)\right)$, we express the resulting SODEs for $\{n_{s,\V{i}}(t)\}$ as
\begin{equation}
\label{SODE_diff_only}
\ddt n_s=D_s\dlapl n_s+\sqrt{\frac{2D_s}{\D{V}}}\ddivg\left(\sqrt{\tilde{n}_s}\mathcal{W}_s\right),
\end{equation}
with the understanding that $\dlapl$ denotes the standard $(2d+1)$-point discrete Laplacian operator, and $\ddivg$ denotes a discrete divergence operator.

\subsection{\label{subsec_spatial_discretization_react_diff}Reaction-Diffusion System}

By combining Eqs.~\eqref{dnsdt} and \eqref{SODE_diff_only} we obtain the spatial discretization of the reaction-diffusion FHD equations as a system of Ito SODEs,  
\begin{equation}
\label{RD_SODE}
\ddt n_s=D_s\dlapl n_s+\sqrt{\frac{2D_s}{\D{V}}}\ddivg\left(\sqrt{\tilde{n}_s}\mathcal{W}_s\right)
+\sum_{r=1}^{N_\mathrm{r}} \frac{\nu_{sr}\mathcal{P}\left(a_r(\V{n})\D{V} dt\right)}{\D{V} dt}.
\end{equation}
In Eq.~\eqref{RD_SPDE_CLE} fluctuations in the reaction rate are modeled as GWN, while in Eq.~\eqref{RD_SODE} we assume Poisson fluctuations.
Since the latter fluctuations are consistent with discrete nature of reactions based on the CME, the description in Eq.~\eqref{RD_SODE} is physically more accurate. 
In fact, it has been shown that the CLE description can give physically incorrect results since it is \emph{not} consistent with a Gibbs--Boltzmann or Einstein equilibrium distribution, even for the case of a single well-mixed cell~\cite{BhattacharjeeBalakrishnanGarciaBellDonev2015}.
As shown above, the inclusion of Poisson fluctuations for reaction, however, requires the notion of a mesoscopic cell and thus can be realized only after the SPDE is spatially discretized.  

The choice of appropriate cell size is a delicate issue for the RDME and FHD descriptions.
An upper bound on the cell size is given by the penetration depth due to the underlying assumption that each cell is homogeneous and reactions occur within a cell.
In fact, there is not only an upper bound of the cell size for a valid description but also a lower bound.
This can be seen by considering the fact that bimolecular reactions would become increasingly infrequent as the cell size decreases~\cite{ErbanChapman2009,Isaacson2009}.
Several criteria for choosing the cell size have been proposed based on physical arguments~\cite{Bernstein2005,IsaacsonPeskin2006,ErbanChapman2009} and mathematical analysis~\cite{KangZhengOthmer2012}.
For a small value of the cell size, corrections in the rate constants of bimolecular reactions have been proposed~\cite{ErbanChapman2009,FangeBergSjobergElf2010,HellanderHellanderPetzold2015}.
However, these corrections do not fix the underlying problem which comes from the fact that reactions are treated as a purely local process with no associated spatial length scale.
In microscopic (particle) models of reaction-diffusion such as the Smoluchowski~\cite{Smoluchowski1917} model or the Doi model~\cite{ErbanChapman2009}, a microscopic reactive distance appears and controls the reaction rate for diffusion-limited reactions.
By introducing a reactive distance into the model, and relaxing the restriction that a reaction should occur among the molecules in the same cell, a modified convergent RDME having well-defined limiting behavior for small cell size can be developed~\cite{Isaacson2013}, and could be combined with our FHD description of diffusion.
\modified{The dependence of stochastic Turing patterns on the grid size has been also investigated~\cite{CaoErban2014}.}

\subsection{\label{subsec_treatments}Maintaining Nonnegative Densities}

The spatially discretized FHD equations~\eqref{SODE_diff_only} or \eqref{RD_SODE} are well defined but suffer from two issues that we now address.
First, the number of molecules in a cell (i.e., $n_{s,\V{i}}\D{V}$) is not an integer.
Second, the cell number density can become negative.
When there are a small number of molecules per cell in the system, the behavior of the FHD description~\eqref{RD_SODE} depends sensitively on the averaging function $\tilde{n}$ and the propensity functions $a_r$, that appear in the multiplicative noise terms.
Hence, we carefully modify the form of $\tilde{n}$ and $a_r$ for negative or very small densities in order to greatly reduce the chances of producing future negative densities.

In Section \ref{subsec_Schlogl_res}, we demonstrate that the arithmetic mean produces more accurate results for the equilibrium distribution than the other Pythagorean means.
Based on the analysis given in Appendix~\ref{appendix_averaging_function}, we use the following modification to the arithmetic mean:
\begin{equation}
\label{avg_type_arithmetic_smoothed}
\tilde{n}(n_1,n_2)=\frac{n_1+n_2}{2}H_0(n_1\D{V})H_0(n_2\D{V}),
\end{equation}
where
\begin{equation}
\label{smoothed_Heaviside_H0}
H_0(x)=
\begin{cases}
0 & (x\le0) \\
x & (0<x<1) \\
1 & (x\ge1)
\end{cases}
\end{equation}
is a smoothed Heaviside function.
The smoothed Heaviside function $H_0$ is introduced to ensure the continuity of 
$\tilde{n}$ at $n_1=0$ or $n_2=0$. 
As explained in Appendix~\ref{appendix_averaging_function}, this averaging function guarantees nonnegativity for the diffusion-only system~\eqref{SODE_diff_only} in the continuous-time description.
In our simulations, we find this modification greatly reduces the occurrence of negative density while closely matching the true equilibrium distribution, noting that in our formulation the stochastic diffusive flux is \emph{continuously} turned off at $n_1 \leq 0$ or $n_2 \leq 0$.
We also note that the smoothing is based on the number of molecules in a cell and if both cells have at least one molecule (i.e., $n_i\D{V}\ge1$), $\tilde{n}$ becomes exactly the arithmetic mean.
As shown in Appendix~\ref{appendix_averaging_function}, the local modification near $n=0$ does not cause any noticeable unphysical behavior for $n_i\D{V}\ge1$.
In Section~\ref{sec_numerics}, we demonstrate that our numerical schemes based on Eq.~\eqref{avg_type_arithmetic_smoothed} work very well even for a small number of molecules per cell.

For the propensity functions $a_r(\V{n})$, we use the following correction to the law of mass action, which is usually included in the RDME description: if the deterministic rate expression contains $n_s^2$ (or $n_s^3$, $\cdots$), replace it by $n_s(n_s-\frac{1}{\D{V}})$ (or $n_s(n_s-\frac{1}{\D{V}})(n_s-\frac{2}{\D{V}})$, $\cdots$).
With this correction, at thermodynamic equilibrium, the mean reaction rate becomes equal to the one calculated from the deterministic rate expression with the mean number density.
This can be seen from the fact that if $n_s {\D{V}}$ follows Poisson statistics with mean $\bar{n}_s {\D{V}}$, $\av{n_s(n_s-\frac{1}{\D{V}})}=\bar{n}_s^2$ and $\av{n_s(n_s-\frac{1}{\D{V}})(n_s-\frac{2}{\D{V}})}=\bar{n}_s^3$.

When reactions are combined with an FHD treatment of diffusion, number densities are no longer restricted to nonnegative integers and special treatment is required to make reaction rates nonnegative and physically sensible for small numbers of molecules. In this work, we evaluate the rate $a_r(\V{n})$ by using continuous-range number densities $\V{n}$ (i.e., without trying to round $\V{n}\D{V}$ to integers) and ensure that each term in the rate of each reaction is nonnegative.
For example, we take the rate expression of the Schl\"ogl model (see Section~\ref{subsec_Schlogl_bg}) to be
\begin{equation}
\label{LMA_Schlogl2}
\textstyle
a(n)= k_1 n^+\left(n-\frac{1}{\D{V}}\right)^+ - k_2 n^+\left(n-\frac{1}{\D{V}}\right)^+\left(n-\frac{2}{\D{V}}\right)^+ + k_3 - k_4 n^+,
\end{equation}
where $n^+=\max(n,0)$.
\modified{We note that more mathematically justified algorithms have been proposed to handle reactions in regards to negative densities using operator splitting and exact solutions of reaction subproblems~\cite{DoeringSargsyanSmereka2005,MoroSchurz2007}; these methods cannot be used to address negative densities due to stochastic diffusive fluxes.}
\section{\label{sec_temporal_integrators}Temporal Integrators}

In this section, we develop temporal integrators for the spatially discretized FHD equation~\eqref{RD_SODE}.
Our goal is twofold.
First, we construct numerical methods that allow for a large time step size even in the presence of fast diffusion.
By treating diffusion implicitly, the severe restriction on time step size can be bypassed.
Second, we construct methods that maintain accuracy even if the time step size is much larger than the diffusive hopping time.
Since it is quite difficult to achieve second-order weak accuracy for general multiplicative noise~\cite{AbdulleVilmartZygalakis2013}, our goal here is to ensure second-order accuracy where possible.
In the limit in which the number of molecules per cell is very large and one can replace random numbers by their means, our schemes reduce to standard second-order schemes for deterministic reaction-diffusion PDEs.
For linearized FHD, our midpoint tau leaping-based schemes are second-order weakly accurate, and all midpoint schemes reproduce at least second-order accurate static correlations, i.e., structure factors.

We build on previous work by some of us~\cite{DonevVandenEijndenGarciaBell2010,DelongGriffithVandenEijndenDonev2013,DelongSunGriffithVandenEijndenDonev2014} and propose two (semi-) implicit schemes as an alternative numerical method to conventional RDME methods.
We mainly consider the case where diffusion is much faster than reaction and molecules on average diffuse more than a cell length per time step (i.e., $2 d D\D{t}\gg\D{x}^2$).
We focus here on \emph{unsplit} schemes that do not rely on operator splitting.
This is because we found that unsplit schemes give notably more accurate structure factors than corresponding split schemes in our case.
In addition, including other transport processes (e.g., advection) and handling boundary conditions~\cite{EinkemmerOstermann2015} to second order is not straightforward for split schemes.

It is convenient to introduce dimensionless numbers, $\alpha$ and $\beta$, which measure how fast diffusion and reaction are relative to the given time step size $\D{t}$, respectively.
For the single-species equation~\eqref{linearized_SPDE}, assuming $\D{x}_1=\cdots=\D{x}_d=\D{x}$, we define
\begin{equation}
\label{alpha_beta}
\alpha=r\D{t},\quad\beta=\frac{D\D{t}}{\D{x}^2},
\end{equation}
where $r$ is the chemical relaxation rate appearing in the linearized equations~\eqref{linearized_SPDE}.
Hence, we can express the well-mixed condition (i.e., the penetration depth $\ell=\sqrt{D/r}\gg\D{x}$) as $\alpha\ll\beta$.
In addition, the numerical stability condition of a scheme can also be given in terms of $\alpha$ and $\beta$.
That is, if reaction and/or diffusion are treated explicitly in a scheme, values of $\alpha$ and/or $\beta$ larger than a stability threshold cause numerical instability.
For $\alpha\ll\beta$, the stability limit is mainly determined by fast diffusion:
\begin{equation}
\label{stab_cond_beta}
\beta\le\frac{1}{2d}\iff\D{t}\le\D{t}_\mathrm{max}\equiv\frac{\D{x}^2}{2d D}.
\end{equation}
Note that the stability limit becomes severe for large diffusion coefficients and small grid spacing and worsens with increasing dimension.

In Section~\ref{subsec_FHD_based_schemes}, we present several numerical schemes for the FHD equation~\eqref{RD_SODE}, including two implicit schemes, and analyze the temporal orders of accuracy for the structure factors using the linearized analysis described in Appendix~\ref{appendix_linearized_eqn_analysis}.
In Section~\ref{subsec_struct_factor_analysis}, we analyze the stochastic accuracy of the numerical schemes for large $\D{t}$ by investigating the structure factor of the one-dimensional Schl\"ogl model at different wavenumbers.
Since analysis for the nonlinear equations is lacking at present, we numerically justify the handling of multiplicative noise in Section~\ref{subsec_Schlogl_res}.

\subsection{\label{subsec_FHD_based_schemes}Schemes}

The simplest method for integrating Eq.~\eqref{RD_SODE} in time is the \emph{Euler--Maruyama tau leaping} (EMTau) scheme,
\begin{equation}
\label{EMTau}
n_s^{k+1} = n_s^k + D_s\D{t}\dlapl n_s^k + \sqrt{\frac{2D_s\D{t}}{\D{V}}}\ddivg{\textstyle\left(\sqrt{\tilde{n}_s^k}W_s^k\right)}
+\sum_{r=1}^{N_\mathrm{r}} \frac{\nu_{sr}\mathcal{P}(a_r^k\D{V}\D{t})}{\D{V}},
\end{equation}
where superscripts denote the point in time at which quantities are evaluated, e.g., $n_{s}^k=n_{s}(k\D{t})$ and $a_{r}^k=a_r(\V{n}(k\D{t}))$, and we have used the compact notation for spatial discretization introduced in Section~\ref{subsec_spatial_discretization_react_diff}. 
Here, $\int_{k\D{t}}^{(k+1)\D{t}} \mathcal{W}(t')dt'$ has been replaced by $\sqrt{\D{t}}W^k$, where $W^k$ denotes a collection of standard random Gaussian variables sampled independently for each species on each grid face at each time step.
That is, the stochastic diffusive flux of species $s$ on face $\V{f}$ at time step $k$ is proportional to $W_{s,\V{f}}^k$.

We also construct numerical schemes where reactions are treated by SSA, which is an exact (exponential) integrator for reactions.
We denote by $\mathfrak{R}_s(\V{n},\tau)$ the (random) change in the number density of species $s$ for a cell with initial state $\V{n}$ obtained from SSA over at time interval $\tau$ (in the absence of diffusion).
We can then write the \emph{Euler--Maruyama SSA} (EM-SSA) scheme as
\begin{equation}
\label{EMSSA}
n_s^{k+1} = n_s^k + D_s\D{t}\dlapl n_s^k + \sqrt{\frac{2D_s\D{t}}{\D{V}}}\ddivg{\textstyle\left(\sqrt{\tilde{n}_s^k}W_s^k\right)}
+\mathfrak{R}_s(\V{n}^k,\D{t}).
\end{equation}

The EMTau scheme is \emph{explicit} in the sense that all terms on the right-hand side of Eq.~\eqref{EMTau} can be evaluated without knowing $n_s^{k+1}$. 
However, a simple analysis shows that the time step size is constrained by a stability condition (for derivation, see Eq.~\eqref{stab_cond_EM})
\begin{equation}
\label{stab_cond_alpha_beta}
\beta+\frac{\alpha}{4d}\le\frac{1}{2d},
\end{equation}
which reduces to condition~\eqref{stab_cond_beta} for $\alpha\ll\beta$.
Since the EM-SSA scheme treats reactions using an exponential integrator, it is only subject to the stability limit~\eqref{stab_cond_beta} without a restriction on $\alpha$.

The stability limit imposed by fast diffusion can be overcome by using standard \emph{implicit} methods such as the second-order implicit midpoint or Crank--Nicolson method, which gives a system of linear equations for $n_s^{k+1}$:
\begin{equation}
\label{CN_EM}
n_s^{k+1} = n_s^k + D_s\D{t}\dlapl \left(\frac{n_s^k+n_s^{k+1}}{2}\right) + \sqrt{\frac{2D_s\D{t}}{\D{V}}}\ddivg{\textstyle\left(\sqrt{\tilde{n}_s^k}W_s^k\right)}
+\sum_{r=1}^{N_\mathrm{r}} \frac{\nu_{sr}\mathcal{P}(a_r^k\D{V}\D{t})}{\D{V}}.
\end{equation}
The linear system~\eqref{CN_EM} can be solved efficiently iteratively using multigrid relaxation~\cite{BriggsHensonMcCormick2000}; for $\beta \lesssim 1$, solving the linear system is not much more expensive than a step of a second-order explicit time stepping scheme.
Note, however, that scheme~\eqref{CN_EM} is only first order accurate overall for reaction-diffusion systems.
Hence, in addition to improved \emph{stability}, it is important to develop higher-order schemes to improve \emph{accuracy}.
Note that this is not as simple as replacing tau leaping in Eq.~\eqref{CN_EM} with SSA as in Eq.~\eqref{EMSSA};
this would still be only first-order accurate even in the deterministic limit.

Here we construct numerical schemes based on the second-order temporal integrators for the linearized equations of FHD developed in Refs.~\cite{DelongGriffithVandenEijndenDonev2013,DelongSunGriffithVandenEijndenDonev2014}.
Those temporal integrators are second-order accurate in the weak sense for additive noise, and are used here as the basis for handling diffusion.
In order to add reactions into diffusion-only schemes, we consider two types of sampling methods, tau leaping and SSA.
For tau leaping, we use the weakly second-order tau leaping method~\cite{HuLiMin2011,AndersonKoyama2012}; a similar two-stage scheme has been originally proposed for the CLE~\cite{AndersonMattingly2011} to achieve second-order weak accuracy.
Here we combine predictor-corrector midpoint schemes proposed in Ref.~\cite{DelongGriffithVandenEijndenDonev2013} (for diffusion) and the second-order tau leaping method (for reaction).
Owing to similar two-stage structures of those schemes, they fit together in a rather natural manner.  
In addition, since the resulting schemes still fit the framework of the implicit-explicit algorithms analyzed in Ref.~\cite{DelongGriffithVandenEijndenDonev2013}, they are second-order weakly accurate for additive noise and an additional order of accuracy (i.e., third order) is gained for the structure factor.

We also develop midpoint schemes that use SSA instead of tau leaping for reactions.
Unlike tau leaping-based schemes, the SSA-based schemes do not suffer from instability even in the presence of rapid reactions.
The use of SSA may also help to prevent the development of negative densities, which is one of the main numerical issues for large fluctuations.
Hence, while SSA-based numerical schemes are computationally more expensive, they work better than tau leaping-based schemes when reactions are fast or when the number of molecules is small.
The SSA-based schemes we propose here belong to a class of exponential Runge--Kutta schemes, and we construct them to ensure second-order deterministic accuracy, as well as second-order accuracy for the structure factor; a detailed analysis of their weak accuracy is at present missing even for linearized FHD.

\subsubsection{\label{subsubsec_ExMid}Explicit Midpoint Schemes}

As a prelude to constructing two-stage implicit methods, we first consider improving the accuracy of the explicit EMTau scheme~\eqref{EMTau} by using an explicit two-stage Runge--Kutta (predictor-corrector) approach. 
By combining the explicit midpoint predictor-corrector scheme from Refs.~\cite{DelongGriffithVandenEijndenDonev2013,DelongSunGriffithVandenEijndenDonev2014} (for diffusion) and the midpoint tau leaping scheme from Refs.~\cite{HuLiMin2011,AndersonKoyama2012} (for reaction), we obtain the \emph{explicit midpoint tau leaping} (ExMidTau) scheme:
\begin{subequations}
\label{ExMidTau}
\begin{align}
&n_s^\star = n_s^k + \frac{D_s\D{t}}{2}\dlapl n_s^k 
+\sqrt{\frac{D_s\D{t}}{\D{V}}} \ddivg\left(\textstyle\sqrt{\tilde{n}_s^k}W_s^\paren{1}\right)
+\sum_{r=1}^{N_\mathrm{r}} \frac{\nu_{sr}\mathcal{P}^\paren{1}(a_r^k\D{V}\D{t}/2)}{\D{V}},\\
&n_s^{k+1} = n_s^k + D_s\D{t}\dlapl n_s^\star
+\sqrt{\frac{D_s\D{t}}{\D{V}}} \ddivg\left(\textstyle\sqrt{\tilde{n}_s^k}W_s^\paren{1}\right)
+\sqrt{\frac{D_s\D{t}}{\D{V}}} \ddivg\left(\textstyle\sqrt{\tilde{n}_s^\bullet}W_s^\paren{2}\right)\\
&\quad +\sum_{r=1}^{N_\mathrm{r}} \frac{\nu_{sr}\mathcal{P}^\paren{1}(a_r^k\D{V}\D{t}/2)}{\D{V}}
+\sum_{r=1}^{N_\mathrm{r}} \frac{\nu_{sr}\mathcal{P}^\paren{2}\left((2a_r^\star-a_r^k)^+\D{V}\D{t}/2\right)}{\D{V}},\notag
\end{align}
\end{subequations}
where the superscripts $\paren{1}$ and $\paren{2}$ indicate that the terms correspond to the first and second half of the time step, respectively.
That is, $\mathcal{P}^\paren{1}$ (and similarly for $W^\paren{1}$ and other random increments) denotes the \emph{same} random number in both predictor and corrector stages, and is only sampled once per time step.
Following Refs.~\cite{HuLiMin2011,AndersonKoyama2012}, the mean reaction rate for the second half step is corrected to $(2a_r^\star-a_r^k)^+$, where $a^+=\max(a,0)$.

For the magnitude of the stochastic diffusive fluxes over the second half of the time step, we consider the following three options for the face average value $\tilde{n}_s^\bullet$:
\begin{subequations}
\label{midpoint_stoch_flux_type}
\begin{align}
\label{midpoint_stoch_flux_type1}
&\tilde{n}_s^\bullet= \tilde{n}\left({n}_s^k\right),\\
\label{midpoint_stoch_flux_type2}
&\tilde{n}_s^\bullet= \tilde{n}\left(n_s^\star\right),\\
\label{midpoint_stoch_flux_type3}
&\tilde{n}_s^\bullet= \tilde{n}\left((2n_s^\star-n_s^k)^+\right). 
\end{align}
\end{subequations}
While all options are consistent with the Ito interpretation, the effect of this choice on accuracy requires a nonlinear analysis that is not available at present.
The option~\eqref{midpoint_stoch_flux_type2} is used in Ref.~\cite{DelongSunGriffithVandenEijndenDonev2014} and shown to lead to second-order weak accuracy for FHD equations linearized around a time-dependent macroscopic state.
The option~\eqref{midpoint_stoch_flux_type3} is inspired by the midpoint tau leaping scheme~\cite{HuLiMin2011,AndersonKoyama2012}.
However, it does not actually lead to second-order weak accuracy for multiplicative noise because the fluctuating diffusion equation does not have the simple noise structure that the CLE has~\cite{AndersonMattingly2011}.
For all our simulations, we use option~\eqref{midpoint_stoch_flux_type3}, as justified by numerical results in Section~\ref{subsec_Schlogl_res}.

The reactions can also be treated using SSA, to give the \emph{explicit midpoint SSA} (ExMidSSA) scheme
\begin{subequations}
\label{ExMidSSA}
\begin{align}
\label{ExMidSSA1}
&n_s^\diamond = n_s^k + \frac{D_s\D{t}}{2}\dlapl n_s^k 
+\sqrt{\frac{D_s\D{t}}{\D{V}}} \ddivg\left(\textstyle\sqrt{\tilde{n}_s^k}W_s^\paren{1}\right),\\
\label{ExMidSSA2}
&n_s^\star = n_s^\diamond + \mathfrak{R}_s^\paren{1}\left(\V{n}^\diamond,\frac{\D{t}}{2}\right),\\
\label{ExMidSSA3}
&n_s^{k+1} = n_s^k + D_s\D{t}\dlapl n_s^\star
+\sqrt{\frac{D_s\D{t}}{\D{V}}} \ddivg\left(\textstyle\sqrt{\tilde{n}_s^k}W_s^\paren{1}\right)
+\sqrt{\frac{D_s\D{t}}{\D{V}}} \ddivg\left(\textstyle\sqrt{\tilde{n}_s^\bullet}W_s^\paren{2}\right)\\
&\quad +\mathfrak{R}_s^\paren{1}\left(\V{n}^\diamond,\frac{\D{t}}{2}\right)
+\mathfrak{R}_s^\paren{2}\left(\V{n}^\star,\frac{\D{t}}{2}\right). \notag
\end{align}
\end{subequations}
Here the predictor stage~\eqref{ExMidSSA1}+\eqref{ExMidSSA2} is a split reaction-diffusion step, but the corrector is not split.
Note that two $\mathfrak{R}^\paren{1}$ appearing in Eqs.~\eqref{ExMidSSA2} and \eqref{ExMidSSA3} are the same random increment computed using SSA.
In other words, the SSA algorithm is called once for each half of the time step; this has the same computational cost as calling SSA once to compute $\mathfrak{R}_s\left(\V{n}^k,\D{t}\right)$ in the EM-SSA scheme~\eqref{EMSSA}.

Since both ExMidTau and ExMidSSA schemes treat diffusion explicitly, they are subject to stability limits.
The ExMidTau scheme has the same stability limit~\eqref{stab_cond_alpha_beta} as the EMTau scheme, whereas the ExMidSSA scheme is subject to the same limit~\eqref{stab_cond_beta} as the EM-SSA scheme.
The ExMidTau scheme with the option~\eqref{midpoint_stoch_flux_type2} is an instance of the explicit midpoint scheme analyzed in Ref.~\cite{DelongSunGriffithVandenEijndenDonev2014} for weak noise (i.e., linearized FHD), and therefore  achieves second-order weak accuracy for linearized FHD and gives third-order accurate equilibrium structure factors. 
On the other hand, the ExMidSSA scheme gives only second-order accurate structure factors.

\subsubsection{\label{subsubsec_ImMid}Implicit Midpoint Schemes}

Here we present two implicit midpoint schemes, where diffusion is treated implicitly based on the implicit midpoint predictor-corrector scheme~\cite{DelongGriffithVandenEijndenDonev2013,DelongSunGriffithVandenEijndenDonev2014}.
By treating reactions using the second-order midpoint tau leaping scheme~\cite{HuLiMin2011,AndersonKoyama2012}, we obtain the \emph{implicit midpoint tau leaping scheme} (ImMidTau) scheme:
\begin{subequations}
\label{ImMidTau}
\begin{align}
\label{ImMidTau_BE_predictor}
&n_s^\star = n_s^k + \frac{D_s\D{t}}{2}\dlapl n_s^\star
+\sqrt{\frac{D_s\D{t}}{\D{V}}}\ddivg\left(\textstyle\sqrt{\tilde{n}_s^k}W_s^\paren{1}\right)
+\sum_{r=1}^{N_\mathrm{r}} \frac{\nu_{sr}\mathcal{P}^\paren{1}(a_r^k\D{V}\D{t}/2)}{\D{V}},\\
&n_s^{k+1} = n_s^k + D_s\D{t}\dlapl \left(\frac{n_s^k+n_s^{k+1}}{2}\right)
+\sqrt{\frac{D_s\D{t}}{\D{V}}} \ddivg\left(\textstyle\sqrt{\tilde{n}_s^k}W_s^\paren{1}\right)
+\sqrt{\frac{D_s\D{t}}{\D{V}}} \ddivg\left(\textstyle\sqrt{\tilde{n}_s^\bullet}W_s^\paren{2}\right)\notag\\
&\quad +\sum_{r=1}^{N_\mathrm{r}} \frac{\nu_{sr}\mathcal{P}^\paren{1}(a_r^k\D{V}\D{t}/2)}{\D{V}}
+\sum_{r=1}^{N_\mathrm{r}} \frac{\nu_{sr}\mathcal{P}^\paren{2}((2a_r^\star-a_r^k)^+\D{V}\D{t}/2)}{\D{V}},
\end{align}
\end{subequations}
where the three options for $\tilde{n}_s^\bullet$ are given in Eq.~\eqref{midpoint_stoch_flux_type}. 
When SSA is used for the reactions, we obtain the \emph{implicit midpoint SSA scheme} (ImMidSSA) scheme,
\begin{subequations}
\label{ImMidSSA}
\begin{align}
&n_s^\star = n_s^k + \frac{D_s\D{t}}{2}\dlapl n_s^\star
+\sqrt{\frac{D_s\D{t}}{\D{V}}} \ddivg\left(\textstyle\sqrt{\tilde{n}_s^k}W_s^\paren{1}\right),\\
&n_s^{k+1} = n_s^k + D_s\D{t}\dlapl \left(\frac{n_s^k+n_s^{k+1}}{2}\right)
+\sqrt{\frac{D_s\D{t}}{\D{V}}} \ddivg\left(\textstyle\sqrt{\tilde{n}_s^k}W_s^\paren{1}\right)
+\sqrt{\frac{D_s\D{t}}{\D{V}}} \ddivg\left(\textstyle\sqrt{\tilde{n}_s^\bullet}W_s^\paren{2}\right)\notag\\
&\quad+\mathfrak{R}_s\left(\V{n}^\star,\D{t}\right).
\end{align}
\end{subequations}
In the corrector stage, both schemes treat diffusion using the Crank--Nicolson method since this gives the most accurate structure factors for diffusion-only systems~\cite{DelongGriffithVandenEijndenDonev2013}.
For the predictor step to the midpoint, we have chosen to use backward Euler for diffusion because this was found to be optimal using the  structure factor analysis discussed in more detail in Section~\ref{subsec_struct_factor_analysis}.

We point out again that the difference in how reactions are included in the ImMidTau and ImMidSSA schemes stems from the fact that SSA is an exponential integrator whereas the midpoint tau leaping method is only a second-order integrator.
This difference must be taken into account when analyzing the accuracy of SSA-based schemes both in the deterministic limit and in structure factor analysis. 
Like the explicit midpoint schemes in Section~\ref{subsubsec_ExMid}, for linearized FHD, the ImMidTau scheme is second-order weakly accurate and gives a third-order accurate structure factor, whereas the ImMidSSA scheme gives only a second-order accurate structure factor.
Since diffusion is treated implicitly in both schemes, they are \emph{not} subject to a stability limit depending on $\beta$. 
However, due to the explicit treatment of reactions, the ImMidTau is subject to the stability condition $\alpha\le 2$.
The ImMidSSA scheme is unconditionally linearly stable and has no stability restrictions on $\alpha$ and $\beta$ but can be considerably more expensive for systems with large numbers of molecules.

A key element of this work that distinguishes
it from our previous work based on CLE \cite{BhattacharjeeBalakrishnanGarciaBellDonev2015} is that here we replaced the GWN in the chemical noise with Poisson noise, and used a weakly second-order tau leaping method~\cite{AndersonKoyama2012,HuLiMin2011} to account for the non-Gaussian nature of the chemical fluctuations.
It is important to note that Poisson noise does not have a continuous range limit, i.e., the Poisson distribution remains integer-valued even as the number of molecules per cell becomes very large.
Although it is tempting to replace the Poisson distribution with a Gaussian distribution, this changes the large deviation functional and therefore we recommend using tau leaping even in the case of weak fluctuations; we note that sampling from a Poisson distribution can be done with a cost essentially independent of the mean using well-designed rejection Monte Carlo methods.
Because of the use of Poisson variates, which cannot be split into a mean and a fluctuation like a Gaussian variate can, there is no strict ``deterministic limit'' for our FHD discretizations.
While the handling of diffusion degenerates to a standard second-order deterministic scheme in the absence of the noise, the chemical noise is always present and increments or decrements the number of molecules by integer numbers.

\subsection{\label{subsec_struct_factor_analysis}Structure Factor Analysis}

Analyzing the accuracy of temporal integrators for stochastic differential equations is notably nontrivial, especially if driven by multiplicative noise. 
As mentioned above, because of the multiplicative noise, all of our midpoint schemes are formally only first-order weakly accurate. 
However, traditional weak-order accuracy is not the most important goal in FHD simulations. 
As first argued in Ref.~\cite{DonevVandenEijndenGarciaBell2010} and then elaborated in Refs.~\cite{DelongGriffithVandenEijndenDonev2013,DelongSunGriffithVandenEijndenDonev2014}, for FHD it is more important to attain \emph{discrete fluctuation-dissipation balance} and higher-order accuracy for the spectrum of the equilibrium fluctuations. 

Here, we analyze the accuracy of our numerical schemes by investigating the structure factor $S(k)$ for the one-dimensional linearized FHD equation~\eqref{linearized_SPDE}.
The analytic expression for $S(k)$ produced by a given scheme can be obtained as a function of $\D{x}$ and $\D{t}$ following the procedure described in Appendix~\ref{appendix_linearized_eqn_analysis}.
Of specific interest to us is how accurately the implicit schemes reproduce $S(k)$ at large wavenumbers corresponding to length scales comparable to $\D{x}$ (i.e., $k \gtrsim (D\D{t})^{-1/2}$) when diffusion is the fastest process, $\beta \gg 1 \gg \alpha$.
This is because incorrect diffusive dynamics at grid scales for $\D{t}\gg\D{x}^2/D$ can lead to gross errors in the magnitude of the fluctuations at large wavenumbers.

Errors in the structure factor arise from two sources: spatial and temporal discretization.
As explained in Appendix~\ref{appendix_linearized_eqn_analysis}, the predominant contribution of spatial discretization is to replace $-k^2$ with the symbol of the standard discrete Laplacian.
In one dimension, this simply amounts to replacing $k$ in the continuum expressions with the modified wavenumber $\tilde{k}$ defined by
\begin{equation}
\label{eq_k_tilde_def}
\tilde{k}=\frac{\sin\left(\frac{k\D{x}}{2}\right)}{\frac{\D{x}}{2}}.
\end{equation}
Note that exactly the same expression applies to the RDME, where diffusion is simulated by lattice hops.
In order to focus our attention on temporal integration errors, we will plot discrete structure factors as a function of $\tilde{k}$ instead of $k$, which effectively removes the spatial errors.

\begin{figure}
\includegraphics[width=0.6\linewidth]{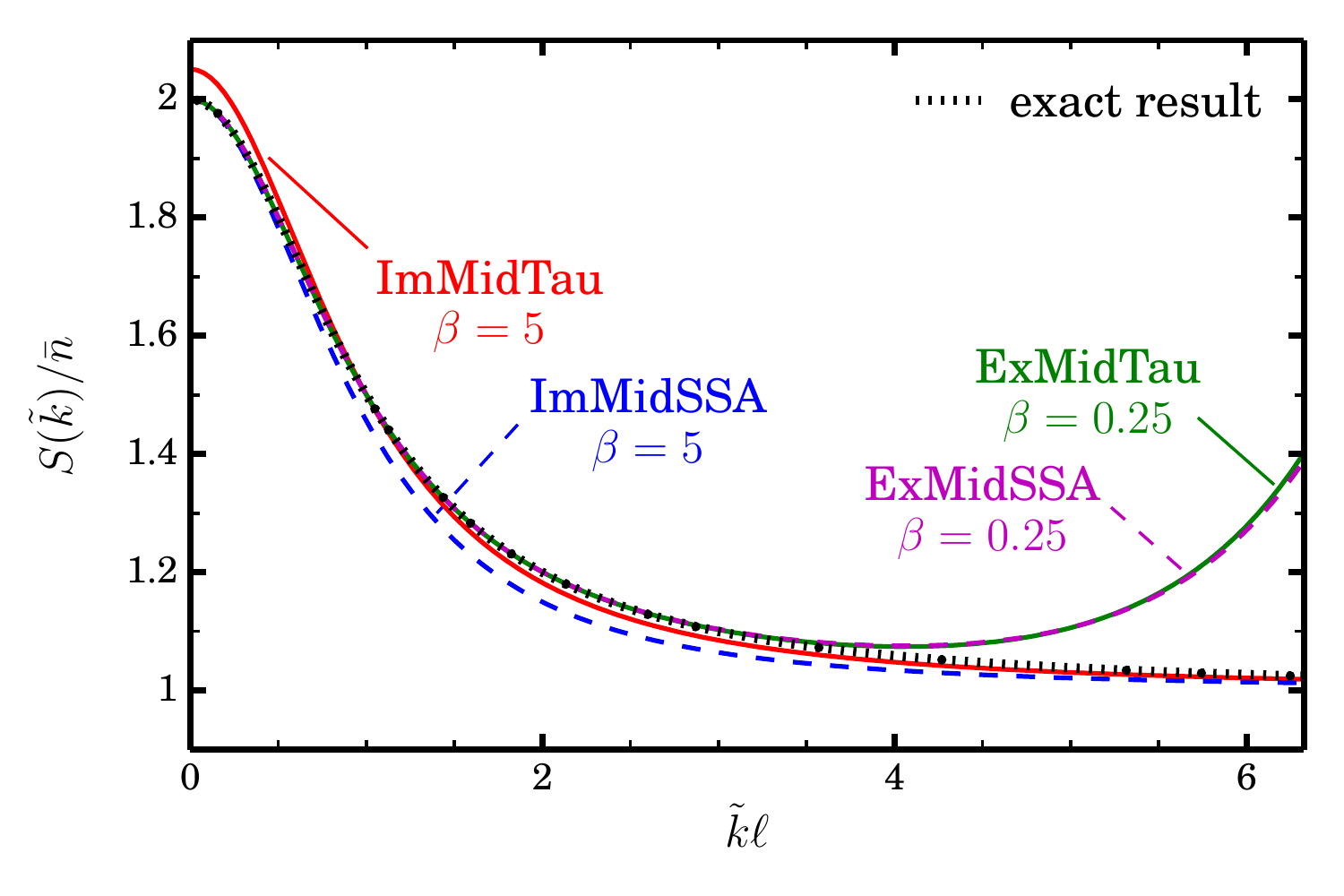}
\caption{\label{fig_Sk_analysis_FHD}
Discrete structure factors $S(\tilde{k})$ for the ExMidTau, ImMidTau, ExMidSSA, and ImMidSSA schemes
for the one-dimensional linearized FHD equation~\eqref{linearized_SPDE} with $\bar{\Gamma}/\bar{n} r=2$ (e.g., an out-of-equilibrium monostable Schl\"ogl model).
Note that different values of $\beta$ are chosen for the explicit schemes ($\beta=0.25$) and the implicit schemes ($\beta=5$) and $\alpha$ is chosen as $\alpha=0.1\beta$.
The exact continuum result~\eqref{Sk_exact} is depicted by the dotted line.}
\end{figure}

Figure~\ref{fig_Sk_analysis_FHD} illustrates how $S(k)$ deviates from the exact result~\eqref{Sk_exact} at different wavenumbers for large $\D{t}$.
We compare $S(k)$ obtained from the four schemes by using values $\alpha=0.025$, $\beta=10 \alpha=0.25$ for the explicit schemes, and $\alpha=0.5$ and $\beta=10\alpha=5$ for the implicit schemes.
Note that these values correspond to a case where the time step size is chosen as half of the stability limit $\D{t}_\mathrm{max}$ for the explicit schemes and, it is increased by factor of 20 ($\D{t}=10\D{t}_\mathrm{max}$) for the implicit schemes.
As described below, the accuracy at diffusion-dominated scales $k \ell \gg 1$ and reaction-dominated scales $k \ell \ll 1$ largely depend on how diffusion (i.e., explicit or implicit) and reaction (i.e., tau leaping or SSA) are treated, respectively.

As the time step size approaches the diffusive stability limit, $\beta\rightarrow 1/2$ in one dimension, the explicit schemes become inaccurate and eventually numerically unstable at the largest wavenumbers, as seen in the figure for $\beta=1/4$.  
Due to the small values of $\alpha$, both schemes give accurate results for reaction-dominated scales.
Note, however, that the SSA-based schemes give exact $S(0)$ and they are in general more accurate at reaction-dominated scales than the corresponding tau leaping-based schemes.
Hence, we see that, for $\beta \gg \alpha$, the accuracy of the explicit schemes is largely affected by numerical instability arising from diffusion.

On the other hand, both implicit schemes give fairly good $S(k)$ in the overall range of $k$ even though $\D{t}$ is twenty times larger and $\beta=5$.
As expected, the ImMidSSA scheme is more accurate for reaction-dominated scales $k\ell\ll 1$, for which the ImMidTau scheme shows some errors because of the relatively large value of $\alpha=0.5$.
However, for intermediate scales $k \ell \sim 1$, the ImMidTau scheme is more accurate because it attains third-order accuracy for static covariances.

At diffusion-dominated scales $k \ell \gg 1$, both implicit schemes give accurate results.
This is not accidental, for we have selected these schemes from a family of schemes parameterized in Ref.~\cite{DelongGriffithVandenEijndenDonev2013} exactly for this reason.
Specifically, the treatment of diffusion in both schemes is based on the implicit midpoint scheme (Crank--Nicolson), which gives the exact $S(k)$ in the absence of reactions~\cite{DelongGriffithVandenEijndenDonev2013}.
In addition, reaction is incorporated in a way that maintains fluctuation-dissipation balance for $k \ell \gg 1$ even for relatively large values of $\alpha$.
For small $\alpha$, the time integration error of the ImMidTau scheme for the structure factor at the maximum wavenumber $k_\mathrm{max}\ell=\pi\sqrt{\beta/\alpha}$ is estimated as
\begin{equation}
\label{SkSktilde}
\frac{S(k_\mathrm{max})-S_0(k_\mathrm{max})}{S_0(k_\mathrm{max})}\approx-\frac{\beta}{2(1+2\beta)^2}\alpha^2,
\end{equation}
where $S_0(k)=\lim_{\D{t}\rightarrow0}S(k)$ is the structure factor in the absence of temporal integration errors (see Appendix~\ref{appendix_linearized_eqn_analysis}).
Hence, for a given value of $\alpha$, $S(k)$ gives accurate results at $k\ell\gg 1$ for large $\beta$.
For the ImMidSSA scheme, $\beta$ in the numerator is replaced by $\beta+(1+2\beta)(\frac{\bar{\Gamma}}{\bar{n}r}-1)$ and a similar stable behavior for large $\beta$ is observed.
By expanding $S(k_\mathrm{max})-S_0(k_\mathrm{max})$ for small $\D{t}$, we also see that the error is $O(\alpha^2\beta)=O(\D{t}^3)$ for the ImMidTau scheme, whereas it is $O(\D{t}^2)$ for the ImMidSSA except at thermodynamic equilibrium (i.e., except when $\bar{\Gamma}=\bar{n}r$), where it is third-order accurate.
\section{\label{sec_numerics}Numerical Results}

We perform numerical simulations for the following three stochastic reaction-diffusion systems.
In Section~\ref{subsec_Schlogl_res}, we use the equilibrium Schl\"ogl model in one, two, and three dimensions to validate our numerical methods.
The analysis in Section~\ref{subsec_struct_factor_analysis} assumed additive noise, reflecting a large number of molecules per cell.
Here we present numerical results demonstrating that the methodology continues to work when there are a small number of molecules per cell and the effects of multiplicative noise are significant.
In particular, we show that our numerical methods, including the modified arithmetic-mean averaging function discussed in Section~\ref{subsec_treatments},
accurately reproduce the Poisson statistics that characterize the thermodynamic equilibrium distribution.
In Sections~\ref{subsec_BPM} and \ref{subsec_Lemarchand_AB}, we study the effects of fluctuations on chemical pattern formation.
In Section~\ref{subsec_BPM}, we test our numerical methods on a time-dependent problem: two-dimensional Turing-like pattern formation in the three-species BPM model~\cite{BarasPearsonMalekMansour1990,BarasMalekMansourPearson1996}.
We investigate how accurately both time-transient and steady state behavior are captured for the ImMidTau and ImMidSSA schemes when a large time step size is used.
We consider the case where the populations of chemical species have different orders of magnitude, which is a frequently encountered situation where a conventional RDME-based method may not work efficiently.
We demonstrate that the ImMidTau scheme scales very well with an increasing number of molecules per cell so that even the deterministic limit of vanishing fluctuations can be explored.
In Section~\ref{subsec_Lemarchand_AB}, we demonstrate the scalability to large systems and computational efficiency
of our FHD approach by presenting a three-dimensional numerical simulation of chemical front propagation in a two-species model~\cite{LemarchandNowakowski2011}.

As a reference method for comparison, we use an RDME-based method, as proposed in Appendix~\ref{appendix_RDME_based_schemes}, which is constructed via a standard operator splitting technique by combining multinomial diffusion sampling ~\cite{BalterTartakovsky2011} and SSA.
Such a split scheme is notably more efficient than ISSA when there are a large number of molecules per cell, and becomes an exact sampling method for the RDME in the limit $\D{t}\rightarrow0$.
This RDME-based scheme works with nonnegative integer populations and reproduces correct fluctuations at thermodynamic equilibrium.
However, diffusion imposes the same restriction~\eqref{stab_cond_beta} on $\D{t}$, and the split scheme produces only a first-order accurate structure factor in general.

In order to set a desired magnitude of fluctuations without changing any parameters \modified{for the macroscopic limit (e.g., penetration depth)}, we introduce a factor $A$, which scales the cell volume $\D{V}=A\D{x}_1\cdots\D{x}_d$.
It can be interpreted as the surface cross section in one dimension and the thickness of a system in two dimensions, and as a rescaling of the number density in three dimensions.
Since the number of molecules in a cell is $n_{s,\V{i}}\D{V}$, the larger $A$ is, the more molecules in a cell there are and the weaker the fluctuations become.
\modified{However, the corresponding macroscopic system is unaffected by the value of $A$.}

\subsection{\label{subsec_Schlogl_res}Schl\"ogl Model at Thermodynamic Equilibrium}

In this section, we test the numerical schemes constructed in Sections~\ref{sec_spatial_discretization} and \ref{sec_temporal_integrators} on the Schl\"ogl reaction-diffusion model, first introduced in Section~\ref{subsec_Schlogl_bg}.
Simulation parameters are chosen to correspond to a system in thermodynamic equilibrium, so that the equilibrium fluctuations are Poisson and the structure factor $S(k)=n_\mathrm{eq}$ is constant, both with and without (i.e., diffusion only) chemical reactions.
Specifically, we set the rate constants as $k_1=k_2=k_3=k_4=0.1$ (see Eq.~\eqref{Schlogl}), which gives $n_\mathrm{eq}=1$ and $\alpha=0.2\D{t}$.
We set the diffusion coefficient $D=1$ and the grid spacing to unity for $d=1,2,3$, and thus $\beta=\D{t}$.
We consider the case where the mean number of molecules per cell is 10 by setting $A=10$.

\subsubsection{Continuous-Time FHD Equation}

\begin{figure}
\centerfloat
\includegraphics[width=1.2\linewidth]{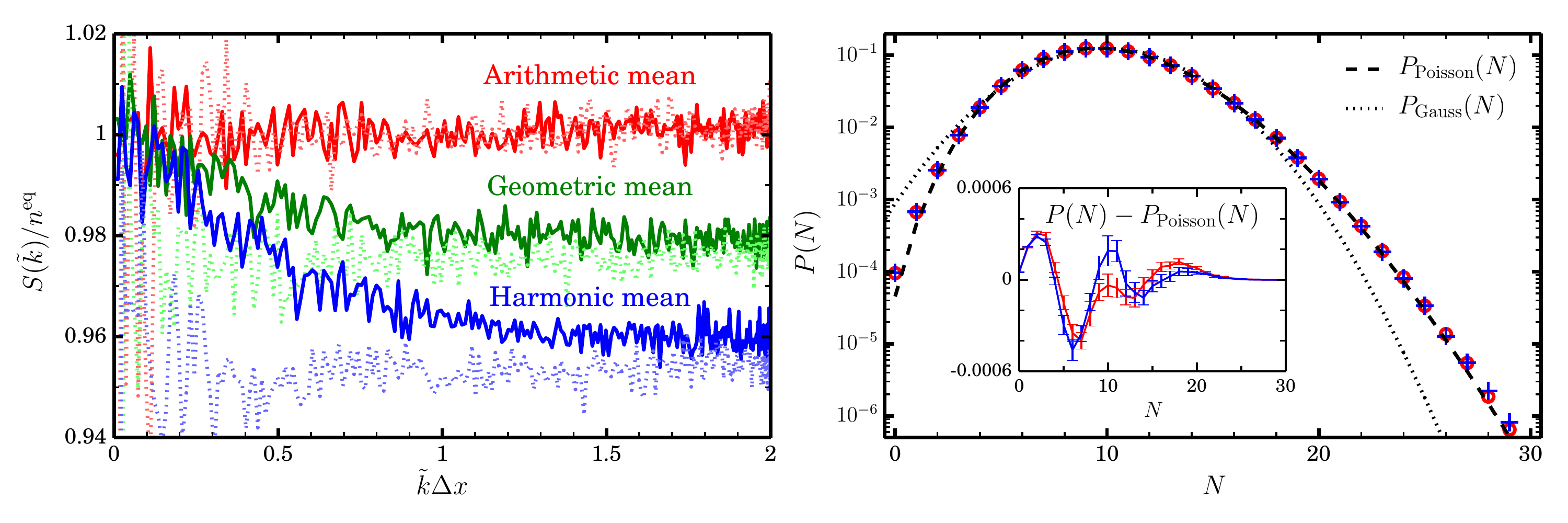}
\caption{\label{fig_Schlogl_Sk_rho}
Static fluctuations of the spatially discretized FHD equation~\eqref{RD_SODE} at thermodynamic equilibrium.
(Left) Structure factors $S(\tilde{k})$ calculated by using different averaging functions $\tilde{n}$.
The solid lines depict the results from the equilibrium Schl\"ogl model ($n_\mathrm{eq}\D{V}=10$), whereas the dotted lines are from the corresponding diffusion-only system.
Note that the exact result for both systems is $S(\tilde{k})=n_\mathrm{eq}$ independent of $\tilde{k}$, which corresponds to Poisson equilibrium fluctuations.
(Right) Empirical histograms $P(N)$ of the number of molecules per cell for the Schl\"ogl model (red circles) and the diffusion-only system (blue crosses), computed using the arithmetic mean averaging function. For comparison, we show the correct Poisson distribution $P_\mathrm{Poisson}(N)$ and its Gaussian approximation $P_\mathrm{Gauss}(N)$.
The inset shows the errors $P(N)-P_\mathrm{Poisson}(N)$ with error bars corresponding to two standard deviations.}
\end{figure}

Prior to evaluating the different temporal integration strategies,
we first focus our attention on the continuous-time discrete-space FHD equation~\eqref{RD_SODE} to establish a baseline for comparison and to evaluate the effectiveness of
the choice of averaging function~\eqref{avg_type_arithmetic_smoothed}.
To eliminate temporal integration errors we use the EMTau scheme~\eqref{EMTau} with a very small time step size $\D{t}=10^{-3}$; results from the other FHD schemes are similar for sufficiently small $\D{t}$.
The left panel of Fig.~\ref{fig_Schlogl_Sk_rho} shows the structure factors $S(k)$ computed for $N_\mathrm{c}=512$ grid cells in one dimension for the arithmetic, geometric, and harmonic mean (AM, GM, HM) averaging functions $\tilde{n}(n_1,n_2)$.
The correct flat spectrum is accurately reproduced by the modified AM averaging function~\eqref{avg_type_arithmetic_smoothed}. 
On the other hand, the GM and HM averaging functions give smaller $S(k)$ at diffusion-dominated scales $k\ell\gg1$.
This can be also observed from the diffusion-only system for all wavenumbers, as theoretically explained in Appendix~\ref{appendix_averaging_function}.  
Henceforth, we use the modified AM averaging function~\eqref{avg_type_arithmetic_smoothed}.

The right panel of Fig.~\ref{fig_Schlogl_Sk_rho} shows that using the AM averaging function, the correct Poisson distribution for the number $N$ of molecules in a cell is accurately reproduced for both reaction-diffusion and diffusion-only systems at thermodynamic equilibrium.
From the equilibrium number density distribution $\rho(n)$, we construct a discrete distribution for integer number of molecules $N$ per cell,
\begin{equation}
\label{disc_P_N}
P(N)=\int_{(N-\frac12)/\D{V}}^{(N+\frac12)/\D{V}}\rho(n)dn,
\end{equation}
and compare it with a Poisson distribution $P_\mathrm{Poisson}(N)$ with mean $n_\mathrm{eq}\D{V}$, as well as a Gaussian distribution $P_\mathrm{Gauss}(N)$ having the same mean and variance as $P_\mathrm{Poisson}(N)$.
The agreement of $P(N)$ and $P_\mathrm{Poisson}(N)$ is remarkable in the sense that FHD was originally proposed to account for only second moments of (small) Gaussian fluctuations.
Since $P_\mathrm{Poisson}(N)$ is significantly different from $P_\mathrm{Gauss}(N)$ for $n^\mathrm{eq}\D{V}=10$, we confirm that our spatially discretized FHD equation~\eqref{RD_SODE} describes (large) Poisson fluctuations faithfully.

\subsubsection{Time Integration Errors}

In order to investigate time integration errors of our numerical schemes, we compare the numerical equilibrium distribution for a given time step size $\D{t}$ with the target Poisson distribution $P_\mathrm{Poisson}(N)$ by using the following measures.
First, we compute the Kullback--Leibler (KL) divergence (distance),
\begin{equation}
\label{KLdiv}
D_\mathrm{KL}=\sum_{N=0}^\infty P_\mathrm{Poisson}(N)\log\frac{P_\mathrm{Poisson}(N)}{P(N)}.
\end{equation}
Second, we compute the probability of negative number densities,
\begin{equation}
\label{Pneg}
P_\mathrm{neg}=\int_{-\infty}^0\rho(n)dn.
\end{equation}
Third, we compute the correlation coefficient $\zeta$ between neighboring cells,
\begin{equation}
\label{corr_zeta}
\zeta = \frac{\Cov{n_{\V{i}}}{n_{\V{i}\pm\V{e}_j}}}{\Var{n}}.
\end{equation}
Note that all three measures should be zero at thermodynamic equilibrium, as they would be for RDME.

\begin{figure}
\centerfloat
\includegraphics[width=1.2\linewidth]{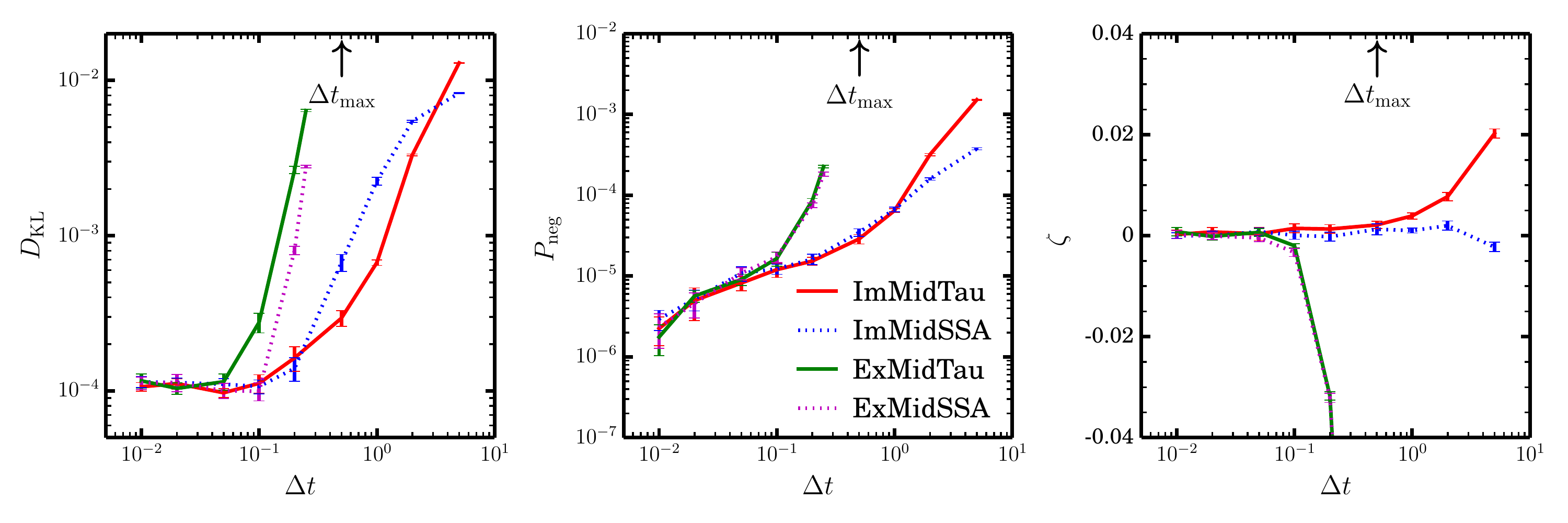}
\caption{\label{fig_Schlogl_time_intg_err}
Deviations from the correct equilibrium distribution $P_\mathrm{Poisson}(N)$ as $\D{t}$ increases for the four midpoint schemes applied to the one-dimensional Schl\"ogl model at thermodynamic equilibrium.  The left panel shows the KL divergence~\eqref{KLdiv}, the middle panel shows the probability of negative density~\eqref{Pneg}, and the right panel shows the correlation coefficient~\eqref{corr_zeta} between neighboring cells.
The red and green solid lines denote the ImMidTau and ExMidTau schemes, respectively, whereas the blue and purple dotted lines denote the ImMidSSA and ExMidSSA schemes, respectively.
The arrows denote the stability limit $\D{t}_\mathrm{max}$ of the explicit schemes, see Eq.~\eqref{stab_cond_beta}.
The error bars correspond to two standard deviations.}
\end{figure}

We considered the three options for the stochastic flux amplitude $\tilde{n}_s^\bullet$ in Eq.~\eqref{midpoint_stoch_flux_type}.
For $\zeta$, the three options give similar values within standard errors of estimation.
For $D_\mathrm{KL}$ and $P_\mathrm{neg}$, option~\eqref{midpoint_stoch_flux_type1} gives the largest values (i.e., least accurate) and option~\eqref{midpoint_stoch_flux_type3} the smallest (not shown).
Based on this result, we will adopt \eqref{midpoint_stoch_flux_type3} and use it for all of the simulations.
Figure~\ref{fig_Schlogl_time_intg_err} shows how these measures deviate from zero as $\D{t}$ increases for $N_\mathrm{c}=64$ cells in one dimension for the different schemes.
As expected, for small values of $\D{t}$, all schemes give similar values.
$D_\mathrm{KL}$ converges to a small value, which is consistent with the good agreement between $P(N)$ and $P_\mathrm{Poisson}(N)$ seen in the right panel of Fig.~\ref{fig_Schlogl_Sk_rho}. 
$P_\mathrm{neg}$ is observed to converge to zero as $\D{t} \rightarrow 0$, which demonstrates the effectiveness of the approach described in Section~\ref{subsec_treatments} and agrees with the analysis in Appendix~\ref{appendix_averaging_function}.
Also, no correlation between neighboring cells is observed for small $\D{t}$ within statistical errors, which is consistent with the flat spectrum $S(k)$ shown in the left panel of Fig.~\ref{fig_Schlogl_Sk_rho}.
As $\D{t}$ approaches the explicit stability limit $\D{t}_\mathrm{max}$, rapid worsening is observed in both explicit schemes in all three measures. 
While $P_\mathrm{neg}$ and $\zeta$ behave similarly for both schemes, $D_\mathrm{KL}$ remains small for larger values of $\D{t}$ for the ExMidSSA scheme compared to the ExMidTau scheme. 

For both implicit schemes, it can be clearly seen that not only is the diffusion instability bypassed but also the accuracy is well maintained for large $\D{t}$.
For comparable accuracy, an order of magnitude larger time step size than the explicit schemes can be chosen.  
The ImMidTau scheme gives smaller $D_\mathrm{KL}$ than the ImMidSSA scheme if $\D{t}$ is smaller than a certain value.
This is consistent with the observation that the former scheme has a higher temporal order of accuracy in $S(k)$. 
However, due to inaccurate handling of reactions by tau leaping for large $\D{t}$, the ImMidTau scheme eventually gives larger $D_\mathrm{KL}$.
For $P_\mathrm{neg}$ and $\zeta$, similar behavior is observed.

Similar behavior is observed for higher spatial dimensions $d=2$ and $d=3$ (not shown).
However, for a given target accuracy tolerance, we find that a smaller time step size should be chosen, which is inversely proportional to the dimensionality $d$.
This should not come as a surprise since the explicit stability limit $\D{t}_\mathrm{max}\sim 1/d$, see Eq.~\eqref{stab_cond_beta}.
Therefore, we conclude that $\D{t}$ can be chosen an order of magnitude larger for the implicit schemes than for the explicit schemes \emph{independent} of the spatial dimension.
\modified{As mentioned, the computational overhead for solving linear systems can be reduced by an efficient iterative solver.
Using multigrid relaxation~\cite{BriggsHensonMcCormick2000}, the overall computational efficiency gain was roughly estimated to be a factor of 3.
However, this factor largely depends on the problem as well as the implementation, especially on the linear solver used.}

\subsection{\label{subsec_BPM}Turing-like Pattern Formation}

In this section, we investigate pattern formation in the three-species Baras--Pearson--Mansour (BPM) model~\cite{BarasPearsonMalekMansour1990,BarasMalekMansourPearson1996},
\begin{gather}
\begin{aligned}
\mathrm{U}+\mathrm{W} & \stackrel{k_1}{\rightarrow} \mathrm{V}+\mathrm{W}, \quad & 2\mathrm{V} & \underset{k_3}{\stackrel{k_2}{\rightleftharpoons}} \mathrm{W}, \\
\mathrm{U} & \underset{k5}{\stackrel{k_4}{\rightleftharpoons}} \varnothing, \quad & \mathrm{V} & \underset{k7}{\stackrel{k_6}{\rightleftharpoons}} \varnothing.
\end{aligned}
\end{gather}
We choose the rate constants so that the deterministic reaction-only system attains a limit cycle as its stable attractor, and we choose the diffusion coefficients so that a Turing-like pattern  forms in the reaction-diffusion system~\cite{BhattacharjeeBalakrishnanGarciaBellDonev2015}.
Specifically, we set $k_1=k_2=\num{2e-4}$, $k_3=1$, $k_4=\num{3.33e-3}$, $k_5=16.7$, $k_6=\num{3.67e-2}$, $k_7 = 4.44$ and $D_\mathrm{U}=0.1$, $D_\mathrm{V}=D_\mathrm{W}=0.01$.
We note that on the limit cycle, number densities of the three species oscillate in significantly different ranges: $n_\mathrm{U}\in(999,2024)$, $n_\mathrm{V}\in(302,645)$, and $n_\mathrm{W}\in(18.2,83.2)$.
For a physical domain with side lengths $L_x=L_y=32$, we use three spatial resolutions with grid sizes $N_\mathrm{c}=64^2$, $128^2$, and $256^2$ cells.
For the initial number densities, we choose a point on the limit cycle $(n_\mathrm{U}^0,n_\mathrm{V}^0,n_\mathrm{W}^0)=(1686,534,56.4)$ and generate the initial number of molecules of each species $s$ in each cell from a Poisson distribution with mean $n_s^0\D{V}=n_s^0 A\D{x}\D{y}$.
We use our implicit schemes in order to bypass the stiff stability limit imposed by the fast diffusion of $\mathrm{U}$ molecules.
To obtain \emph{reference FHD results} having minimal time integration errors, we use the ImMidTau scheme with $\D{t}=0.1$.
To test the importance of fluctuations, a deterministic version of the ImMidTau scheme is used with $\D{t}=0.1$, with random initial conditions generated from a Poisson distribution corresponding to thickness $A=10$.


\begin{figure}
\centerfloat
\includegraphics[width=1.2\linewidth]{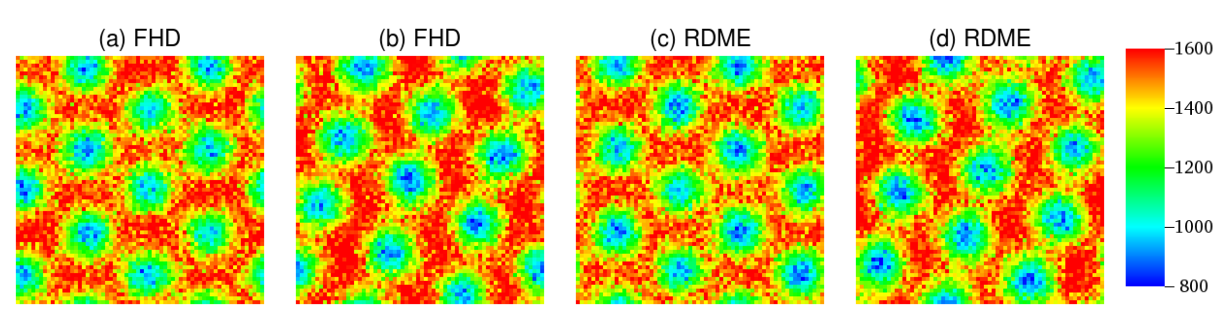}
\caption{\label{fig_BPM_final}
Two types of steady-state Turing-like patterns observed in the two-dimensional BPM model with $64^2$ cells and $A=1$.
Snapshots of $n_\mathrm{U}$ are obtained at $t=\num{5e4}$ from FHD (left two panels) and RDME (right two panels) simulations with a small time step size. Panels~(a) and (c) show a hexagonal structure (with 12 dots), whereas panels~(b) and (d) exhibit a monoclinic structure (with 11 dots).}
\end{figure}

Figure~\ref{fig_BPM_final} shows snapshots of a final Turing pattern formed for $A=1$ and $64^2$ cells.
While the pattern is qualitatively correct, the quantitative behavior of our FHD formulation may be questioned since the mean number of $\mathrm{W}$ molecules in a cell can be as low as $4.5$ at small $t$ in this case.
To confirm the FHD description applies even for relatively small numbers of molecules per cell, we compare the FHD results to reference RDME results obtained using the SSA/2+MN+SSA/2 scheme~\eqref{SSA2MNSSA2} with $\D{t}=0.01$.
We find that the FHD reference simulations are qualitatively very similar to the RDME reference simulations over a wide range of thicknesses $A$, as we illustrate in Fig.~\ref{fig_BPM_final}. 
For our setup, after the initial formation of a disordered pattern of dots with low concentration of $\mathrm{U}$ molecules (blue dots in Fig~\ref{fig_BPM_final}), the dots split and merge and diffuse to eventually form a stable regular pattern; note that the final patterns are nearly periodic lattice structures but their geometry is affected to some extent by the finite size of the domain.
For $A=1$, by $t=\num{5e4}$, almost all samples had formed a steady pattern. Most samples formed a hexagonal (12 dots, see panels~(a) and (c) in Fig~\ref{fig_BPM_final}), and a few formed a monoclinic (11 dots, see panels~(b) and (d) in Fig.~\ref{fig_BPM_final}) lattice of dots, for both FHD and RDME.
Note that while FHD simulations using the ImMidTau scheme are equally efficient independent of $A$, RDME simulations become prohibitively expensive for large $A\gtrsim 100$ due to the very large number of $\mathrm{U}$ molecules (as many as $\num{2e6}A$) in the system. 
For weaker fluctuations, $A=1000$, FHD simulations reveal that the annealing of the lattice defects takes much longer and we see several disordered or defective patterns even at $t=\num{5e4}$ (not shown).
Therefore, not only do fluctuations accelerate the formation of the initial (disordered) pattern, but they also appear to accelerate the annealing of the defects.


\begin{figure}
\centerfloat
\includegraphics[width=1.2\linewidth]{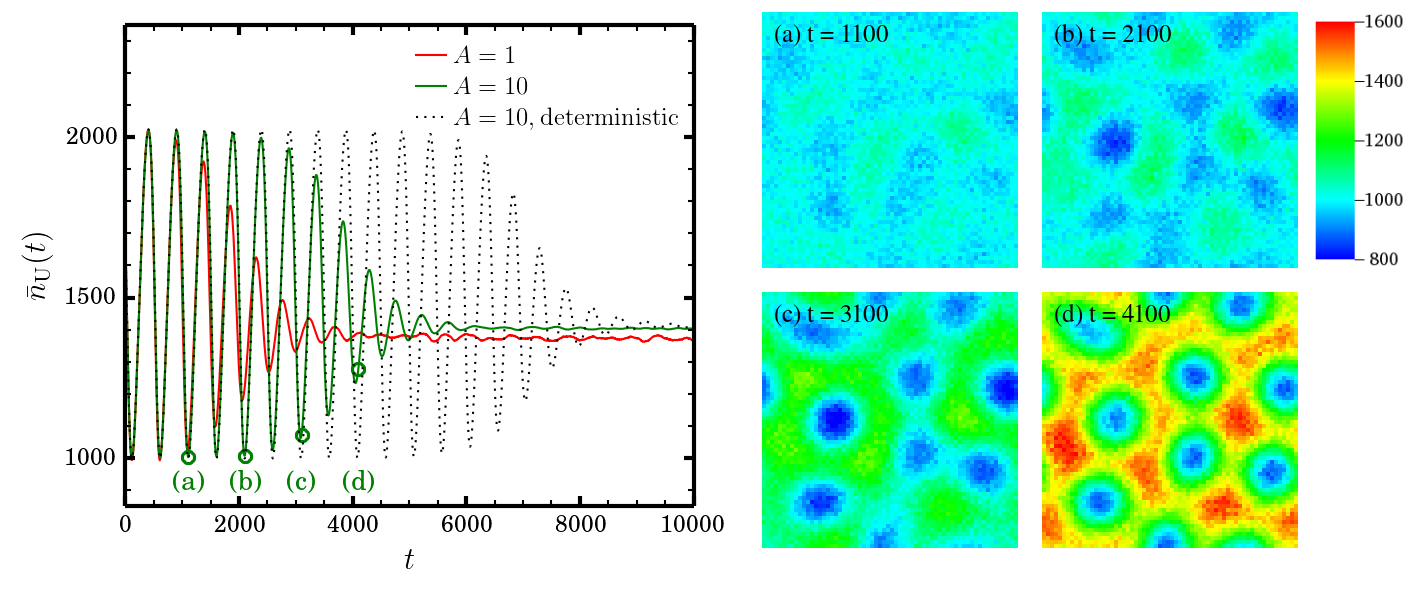}
\caption{\label{fig_BPM_Turing}
Turing-like pattern formation in the two-dimensional BPM model with $64^2$ cells.
(Left)~Spatially-averaged density $\bar{n}_\mathrm{U}(t)$ of species $\mathrm{U}$ for domain thickness $A=1$ and $A=10$ (RDME results), as well as deterministic reaction-diffusion started from random initial conditions corresponding to $A=10$.
(Right)~Snapshots of $n_\mathrm{U}$ for $A=10$ at four different times $t$ at which $\bar{n}_\mathrm{U}(t)$ attains a local minimum, indicated by circles in the left panel.}
\end{figure}

Since the formation of the pattern is driven by an instability, it is itself a random process and a proper quantitative comparison between the different methods requires a careful statistical analysis of an ensemble of trajectories.
In order to capture the time transient behavior of pattern formation, illustrated in the right panel of Fig.~\ref{fig_BPM_Turing}, we calculate the spatially-averaged density $\bar{n}_\mathrm{U}(t)=\frac{1}{N_\mathrm{c}}\sum_{\V{i}}n_{\mathrm{U},\V{i}}(t)$.
In the left panel, we compare sample trajectories of $\bar{n}_\mathrm{U}(t)$ for $A=1$ and $A=10$ for RDME (similar results are obtained for FHD) and for deterministic reaction-diffusion.
While $\bar{n}_\mathrm{U}(t)$ initially oscillates as in the limit cycle, as the Turing-like pattern begins to form, the oscillation amplitude decays and $\bar{n}_\mathrm{U}(t)$ eventually attains a steady value.
By comparing $A=1$ and 10, we see that larger fluctuations facilitate faster pattern formation, as observed in prior work~\cite{BhattacharjeeBalakrishnanGarciaBellDonev2015} by us and others.
By comparing RDME results for thickness $A=10$ with deterministic reaction-diffusion started from the same initial condition, we see that the effect of fluctuations on pattern formation is \emph{not} just due to random initial conditions.


\begin{figure}
\centerfloat
\includegraphics[width=1.2\linewidth]{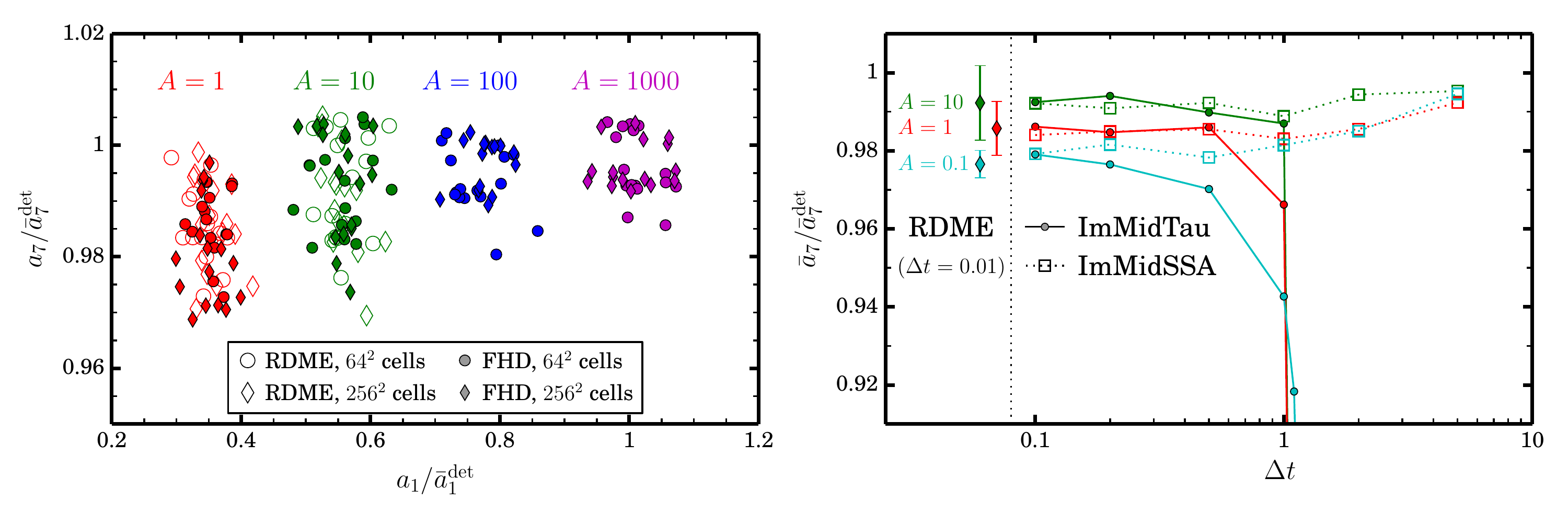}
\caption{\label{fig_BPM_comp}
(Left)~Scatter plots of the decay onset time $a_1$ and the steady spatial average density $a_7$ for several values of the cell thickness $A$ and two grid resolutions.
RDME and FHD results are compared for $A=1$ and $A=10$, whereas only the FHD results (using the ImMidTau scheme) are shown for $A=100$ and $A=1000$ since RDME simulations are prohibitively expensive.
(Right)~Average values $\bar{a}_7$ of $a_7$ over 16 samples for the ImMidTau and ImMidSSA schemes with $64^2$ cells as a function of time step size $\D{t}$, for cell thicknesses $A$ of 0.1, 1, and 10.
For comparison, the RDME results for $\D{t}=0.01$ are shown on the left with error bars corresponding to two standard deviations.
Error bars are omitted for the implicit schemes; they are comparable to the RDME results.
Note that $a_i$ are normalized by the average values $\bar{a}_i^\mathrm{det}$ for deterministic reaction-diffusion started with random initial conditions corresponding to $A=10$.}
\end{figure}

We generate 16 sample trajectories for each set of parameter values, fit each realization of $\bar{n}_\mathrm{U}(t)$ using seven fitting parameters $a_1,\cdots,a_7$ to
\begin{equation}
\bar{n}_\mathrm{U}(t)=\left(1-\tanh\frac{t-a_1}{a_2}\right)\Big(a_3\sin(a_4 t+a_5)+a_6\Big)+a_7,
\end{equation}
and compare the distributions of the fitting parameters.
Note that $a_1$ and $a_7$ correspond to the decay onset time and the steady spatial average density, respectively.
In the left panel of Fig.~\ref{fig_BPM_comp}, we compare the empirical distributions of $(a_1,a_7)$ from the RDME and FHD results for different values of the thickness $A$ and spatial resolutions.
For each value of $A$, we observe that distributions obtained from different methods and/or resolutions coincide.
For $A=1$ and $A=10$, we reconfirm that the RDME approach produces statistically very similar results for three resolutions and the FHD results are statistically indistinguishable from the RDME results.
It is quite remarkable that FHD works even for $A=1$ and $256^2$ cells, which can have as low as 0.3 molecules per cell at small $t$, and this demonstrates the robustness of our treatment for a small number of molecules per cell.
As the magnitude of the fluctuations increases (i.e., as $A$ decreases), the pattern begins to form earlier (i.e., $a_1$ decreases), as already seen in Fig.~\ref{fig_BPM_Turing}, while the steady spatial average density $a_7$ becomes smaller.
In addition, while the variance of $a_1$ does not change significantly as $A$ varies, the variance of $a_7$ becomes larger for smaller $A$. 

Finally, we investigate time integration errors of our implicit schemes for the Turing-like pattern formation.
For the ImMidTau scheme, we increase $\D{t}$ up to the stability limit arising from reactions $\D{t}_\mathrm{max}^\paren{\mathrm{R}}\approx1.3$.
The right panel of Fig.~\ref{fig_BPM_comp} shows the mean values $\bar{a}_7$ of the steady spatial average density $a_7$ over 16 samples versus $\D{t}$ for $64^2$ cells.
While both schemes give similar values to the RDME results for small $\D{t}$, they show different behavior for large $\D{t}$, which also depends on $A$.
As expected, in the ImMidTau scheme, the value of $\bar{a}_7$ rapidly deviates from the RDME result as $\D{t}$ approaches $\D{t}_\mathrm{max}^\paren{R}$, especially if there are few molecules per cell, $A=0.1$ or $A=1$.
On the other hand, in the ImMidSSA scheme, $\D{t}$ can be increased beyond $\D{t}_\mathrm{max}^\paren{R}$, and deviations from the RDME results remain small even for the smallest value of $A$.
Hence, handling reactions by SSA not only removes the reaction stability constraint but also improves the accuracy for a small number of molecules per cell.
However, it should be noted that this improvement comes at a significant computational cost, since the SSA scheme is much more expensive than tau leaping especially as the number of molecules per cell increases.
Therefore, the SSA-based schemes are impractical in the regime of weak fluctuations due to poor scaling.
We discuss some alternatives to SSA that may significantly improve the computational cost for weaker fluctuations in the Conclusions.

\subsection{\label{subsec_Lemarchand_AB}Front Propagation}

\begin{figure}
\centerfloat
\includegraphics[width=1.2\linewidth]{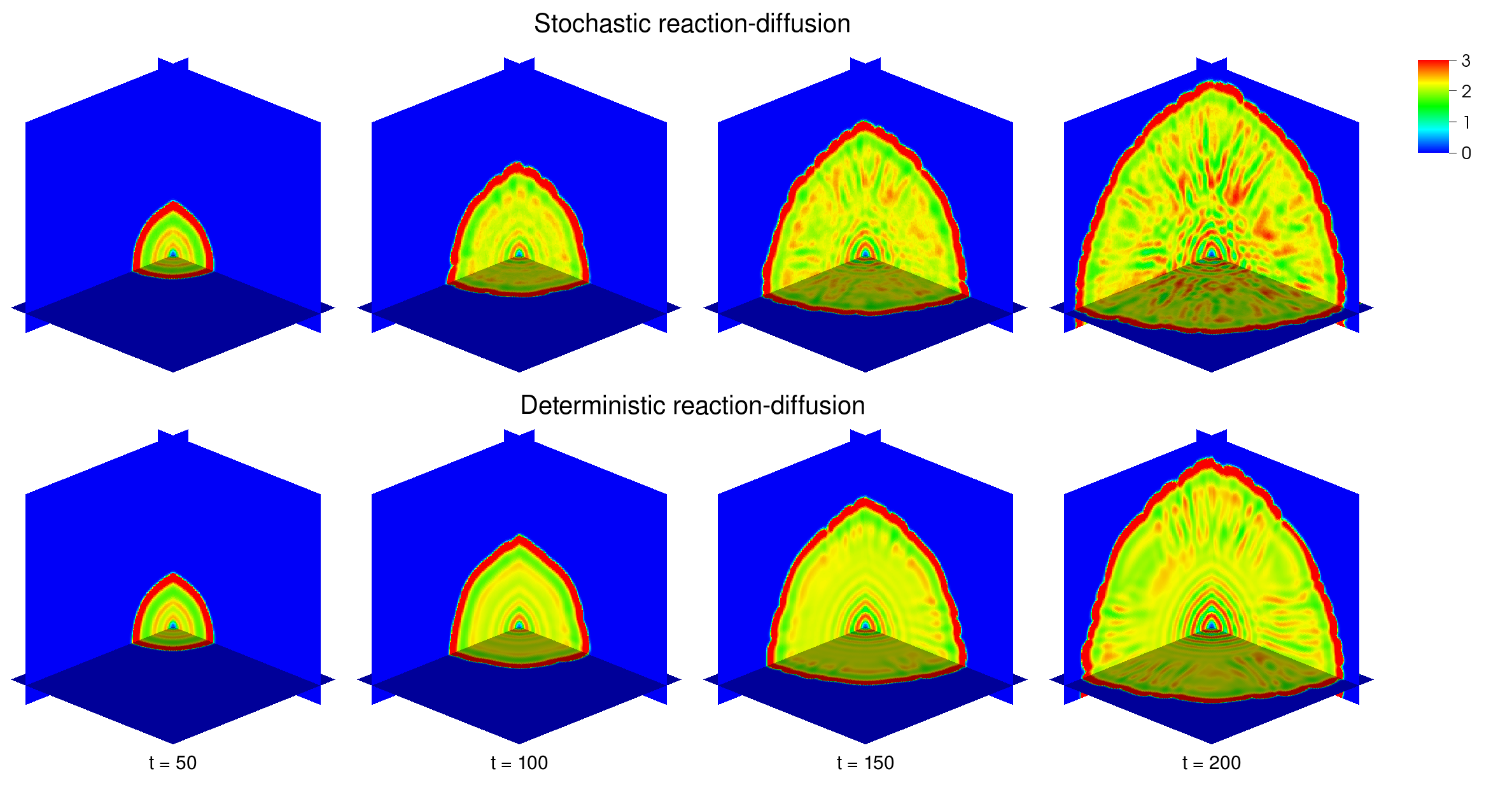}
\caption{\label{fig_AB_bubble_3d}Three-dimensional FHD simulation of front propagation in a two-species stochastic reaction-diffusion model~\eqref{Lemarchand_AB} using the ImMidTau scheme.
The number density of species $\mathrm{A}$ is shown at four different times $t$ for the stochastic reaction-diffusion system (in the top row) and the corresponding deterministic case (in the bottom row).
The same initial conditions with Poisson fluctuations are used in both simulations.}
\end{figure}

As a final example, we simulate three-dimensional front propagation in a two-species stochastic reaction-diffusion system having the following reaction network: 
\begin{equation}
\label{Lemarchand_AB}
\mathrm{A}\stackrel{k_1}{\rightarrow}\varnothing,\quad
2\mathrm{A}+\mathrm{B}\stackrel{k_2}{\rightarrow}3\mathrm{A},\quad
\mathrm{B}\underset{k_4}{\stackrel{k_3}{\rightleftharpoons}}\varnothing.
\end{equation}
This model has been proposed to reproduce axial segmentation in Ref.~\cite{LemarchandNowakowski2011}, where ISSA simulations have been performed for the one-dimensional case.
Following Ref.~\cite{LemarchandNowakowski2011}, we set $k_1=0.4$, $k_2=0.137$, $k_3=0.1$, $k_4=1$ and $D_\mathrm{A}=1$, $D_\mathrm{B}=10$.
For a physical domain with side lengths $L_x=L_y=L_z=512$, we use $256^3$ cells.
To initiate front propagation, we generate initial number densities as follows.
We first assign to each cell ${\V{i}}$ and species $s$ a mean number density
\begin{equation}
n_{s,\V{i}}^0 = n^\paren{1}_s+\frac12\left(1+\tanh\frac{r_{\V{i}}-R}{\xi}\right)\left(n^\paren{2}_s-n^\paren{1}_s\right),
\end{equation}
where $r_{\V{i}}$ is the distance from the cell center to the center of the domain.
This initializes a spherical region of radius $R$ in the first uniform equilibrium state of the model, $(n_\mathrm{A}^\paren{1},n_\mathrm{B}^\paren{1})=(2.16, 1.35)$, while the rest of the domain is initialized in the second uniform equilibrium state, $(n_\mathrm{A}^\paren{2},n_\mathrm{B}^\paren{2})=(0, 10)$, with a smooth transition region between the two states of width $\approx 2\xi$.
Then, as in Section~\ref{subsec_BPM}, we generate the initial number of molecules of each species in each cell from a Poisson distribution with mean $n_{s,\V{i}}^0\D{V}=n_{s,\V{i}}^0 A\D{x}\D{y}\D{z}$.
We simulate the system for parameters $A=1000$, $R=16$, and $\xi=4=2\D{x}$ using the ImMidTau scheme with $\D{t}=0.25$.
For comparison, we also simulate the corresponding deterministic system using a deterministic version of the ImMidTau scheme with the same time step size and (random) initial conditions.
Simulations are performed using a parallel implementation of the algorithm using the BoxLib software framework~\cite{RendlemanBecknerLijewskiCrutchfieldBell2000}.
We emphasize that a corresponding RDME system is too large to simulate with conventional RDME-based methods; while the total number of molecules in the system varies as the front propagates, it is of the order of $10^{12}$ for $A=1000$.

Figure~\ref{fig_AB_bubble_3d} shows the growth of the spherical region as the more stable phase propagates into the less stable phase via a spherical traveling wave.
While the phase boundary having a peak population of species $\mathrm{A}$ propagates, a Turing-like pattern develops behind the wave front; in one dimension this pattern is periodic and more pronounced in the presence of fluctuations~\cite{LemarchandNowakowski2011}.
In two and three dimensions, fluctuations not only enhance the pattern but they also make it disordered, as seen in the figure by comparing the stochastic and deterministic cases.
In addition, the phase boundary becomes more irregular under fluctuations. 
Note that the numerical results for the deterministic case are not perfectly radially symmetric not only due to the noisy initial conditions but also due to grid artifacts introduced by the standard discrete Laplacian, which is not perfectly isotropic~\cite{RamaduguThampiAdhikariSucciAnsumali2013} on length scales compared to the front width (i.e., the penetration depth); one would require an even finer grid to correct for this spatial discretization artifact.
\section{\label{sec_conclusions}Conclusions}

In this work, we have formulated a fluctuating hydrodynamics (FHD) model for reaction-diffusion systems and developed numerical schemes to solve the resulting stochastic ordinary differential equations (SODEs)~\eqref{RD_SODE} for the number densities $n_{s,\V{i}}(t)$ of chemical species in each cell.
We obtained the diffusion part of the SODEs from an FHD description of a microscopic system consisting of molecules undergoing independent Brownian motions, and added reactions in an equivalent manner to the reaction-diffusion master equation (RDME).
We presented two implicit predictor-corrector schemes, the ImMidTau~\eqref{ImMidTau} and ImMidSSA~\eqref{ImMidSSA} schemes, that treat reactions using tau leaping and SSA, respectively.
In these schemes, diffusion is treated implicitly so that the stability limit imposed by fast diffusion can be bypassed and the time step size can be chosen to be significantly larger than the hopping time scale of diffusing molecules.
In addition, two-stage Runge--Kutta temporal integrators are employed to improve the accuracy.
To confirm the validity of our FHD formulation and demonstrate the performance of our numerical schemes, we numerically investigated not only a system at steady state (Schl\"ogl reaction-diffusion model), but also time-dependent two-dimensional Turing-like pattern formation and three-dimensional front propagation.

Based on our analytical and numerical investigation, we conclude that the ImMidTau scheme is an efficient and robust alternative numerical method for reaction-diffusion simulations.
The reason is threefold.
First, the cost of the scheme does not increase for increasing number of molecules per cell (weaker fluctuations).
For small numbers of molecules per cell (large fluctuations), the integer-valued RDME description is more appropriate than the continuous-range FHD description.
However, by using the approach proposed in Section~\ref{subsec_treatments}, we ensured that the FHD description remains robust and gives accurate results even for a small number of molecules per cell, as shown in Section~\ref{subsec_Schlogl_res}.
Hence, as shown in Section~\ref{subsec_BPM}, our numerical methods can efficiently simulate reaction-diffusion systems over a broad range of relative magnitude of the fluctuations.
Second, the scheme allows a significantly larger time step size without degrading accuracy
compared to existing RDME-based numerical methods~\cite{RodriguezKaandorpDobrzynskiBlom2006,LampoudiGillespiePetzold2009,EngblomFermHellanderLotstedt2009,FermHellanderLotstedt2010,DrawertLawsonPetzoldKhammash2010,HellanderLawsonDrawertPetzold2014}, which use a fixed time step size for diffusion that is comparable to the hopping time scale.
In particular, we found that the time step size could be chosen an order of magnitude larger for the implicit schemes than for explicit methods, independent of the spatial dimension.
Lastly, FHD can take advantage of development of efficient parallel algorithms developed for computational fluid dynamics (CFD) that enable effective use of high-performance parallel architectures while providing the framework for treating more complex problems with additional physical phenomena.
This enabled us to perform three-dimensional simulations of chemical front propagation involving as many as $10^{12}$ molecules using the BoxLib CFD software framework~\cite{RendlemanBecknerLijewskiCrutchfieldBell2000}.

The explicit tau leaping methods used here are quite simple to implement but are subject to a stability limit for fast reactions, and can lead to negative densities when fluctuations are large.
While some implicit tau leaping methods have been developed, as an alternative we developed methods that use SSA for reactions.
The ExMidSSA and ImMidSSA methods, however, do not scale as the fluctuations become weaker.
This deficiency can be corrected by replacing SSA by a recently-developed hybrid algorithm termed asynchronous tau leaping~\cite{JdrzejewskiSzmekBlackwell2016} that combines SSA and tau leaping in a dynamic manner by simulating multiple events with asynchronous time steps.
\deleted{A nontrivial question that requires further study is how to control the accuracy of such methods.
In fact, the standard notion of temporal order of accuracy does not apply to asynchronous algorithms since there is no notion of a fixed time step size.}
Future work should develop FHD-based numerical schemes that are accurate and robust even for a small number of molecules per cell and also scale to the deterministic limit efficiently.

One of the advantages of the FHD approach for reaction-diffusion systems is its natural generalization to more complicated and realistic applications.
Chemical reactions of interest usually occur in liquid solution, and often in a dense crowded environment such as the cytoplasm~\cite{AndoSkolnick2010,DixVerkman2008,Skolnick2016}.
It is well-known that Brownian motion of liquid molecules or suspended macromolecules in liquids is dominated by hydrodynamic effects related to viscous dissipation~\cite{DonevFaiVandenEijnden2014,DonevVandenEijnden2014,Skolnick2016}.
This means that the diffusion model commonly used in reaction-diffusion models, including this work, which assumes that reactants are independent non-interacting Brownian walkers diffusing with a constant diffusion coefficient, does not apply in the majority of practical problems of interest. Notably, crowding or steric interactions affect the local hydrodynamic mobility of individual reactants, and hydrodynamic interactions (HI) among the diffusing reactants introduce strong correlations among the diffusive motions of the reacting particles (and also among reactants and passive crowding agents)~\cite{Skolnick2016}.
Excluded volume due to steric repulsion introduces cross-diffusion effects, i.e., coupling between the diffusive fluxes for different species~\cite{SignonNowakowskiLemarchand2016}, as well as concentration-dependent diffusion coefficients. 
Furthermore, it has been observed that cross-diffusion may lead to qualitatively different Turing instabilities~\cite{BiancalaniFanelliPatti2010,FanelliCianciPatti2013,KumarHorsthemke2011}.
Long-ranged contributions of hydrodynamic interactions can be captured by accounting for the advection of concentration fluctuations by the thermal velocity fluctuations, which follow a fluctuating Stokes equation~\cite{DonevFaiVandenEijnden2014,DonevVandenEijnden2014}.
Additional thermodynamic contributions to the diffusive fluxes such as cross-diffusion, barodiffusion and thermodiffusion do not seem straightforward in the RDME but are easily included in our FHD formulation~\cite{BhattacharjeeBalakrishnanGarciaBellDonev2015}.
In future work, we will investigate these hydrodynamic effects on stochastic reaction-diffusion phenomena.

\begin{acknowledgments}
We would like to thank Jonathan Goodman for numerous informative discussions.
This work was supported by the U.S.~Department of Energy, Office of Science, Office of Advanced Scientific Computing Research, Applied Mathematics Program under contract DE-AC02-05CH11231 and Award Number DE-SC0008271.
This research used resources of the National Energy Research Scientific Computing Center, a DOE Office of Science User Facility supported by the Office of Science of the U.S.~Department of Energy under Contract No.~DE-AC02-05CH11231.
\end{acknowledgments}

\appendix

\section{\label{appendix_RDME_based_schemes}RDME-Based Split Scheme}

As a reference algorithm for stochastic reaction-diffusion simulation, we construct a numerical scheme for the RDME through a standard operator splitting technique, as done in a number of prior  works~\cite{EngblomFermHellanderLotstedt2009,FermHellanderLotstedt2010,HellanderLawsonDrawertPetzold2014,ChoiMauryaTartakovskySubramaniam2012}.
This technique allows one to obtain a numerical scheme for the reaction-diffusion system by combining numerical methods for the diffusion-only and reaction-only systems.
Here we combine multinomial diffusion sampling~\cite{BalterTartakovsky2011,LampoudiGillespiePetzold2009} for diffusion and SSA for reaction via Strang splitting~\cite{Strang1968}.

One distinguishing feature of the resulting scheme, compared to exact sampling methods such as ISSA, is the use of a fixed time step size $\D{t}$ for diffusion, which facilitates an efficient numerical simulation if diffusion is fast.
As shown below, while there are several advantages of this scheme over ISSA, the choice of $\D{t}$ is constrained as it is for explicit FHD schemes.

In Section~\ref{subsec_RDME_based_schemes}, after reviewing the multinomial diffusion sampling method, we present an RDME-based split scheme and discuss its advantages and disadvantages.
In Section~\ref{subsec_struct_factor_analysis_RDME}, we present a stochastic accuracy analysis for this scheme.

\subsection{\label{subsec_RDME_based_schemes}Split Scheme}

In the multinomial diffusion (MN) sampling method, the numbers of molecules in each cell after time $\D{t}$ are calculated by sampling how many molecules have moved from a cell to a neighboring cell. Here we follow the simple algorithm described by Balter and Tartakovsky~\cite{BalterTartakovsky2011} and only allow particles to move to nearest-neighbor cells.
More complicated but also more accurate algorithms that allow particles to jump to further than nearest-neighbor cells are described by Lampoudi et al.~\cite{LampoudiGillespiePetzold2009}.
We use the notation introduced in the body of the paper.

By denoting the number of molecules of species $s$ diffusing from cell $\V{i}$ to cell $\V{i}'$ over the time interval $\D{t}$ as $N_{s,\V{i}\rightarrow\V{i}'}$, the change in the number density $n_{s,\V{i}}$ can be expressed in terms of the sum of the inward and outward fluxes, 
\begin{subequations}
\label{MN}
\begin{align}
\label{MN_form1}
n_{s,\V{i}}(t+\D{t}) &= n_{s,\V{i}}(t) 
+\frac{1}{\D{V}}\sum_{j=1}^d \left(
N_{s,\V{i}+\V{e}_j\rightarrow\V{i}}+N_{s,\V{i}-\V{e}_j\rightarrow\V{i}}
-N_{s,\V{i}\rightarrow\V{i}+\V{e}_j}-N_{s,\V{i}\rightarrow\V{i}-\V{e}_j}
\right)\\
\label{MN_form2}
&= n_{s,\V{i}}(t)+\mathfrak{D}_{\V{i}}(n_s(t),{\D{t}}),
\end{align}
\end{subequations}
where $n_s(t)=\{n_{s,\V{i}}(t)\}$.
For each cell $\V{i}$, the outward fluxes $(N_{s,\V{i}\rightarrow\V{i}+\V{e}_1}, N_{s,\V{i}\rightarrow\V{i}-\V{e}_1}, \cdots, N_{s,\V{i}\rightarrow\V{i}+\V{e}_d},$ $ N_{s,\V{i}\rightarrow\V{i}-\V{e}_d}, N_{s,\V{i}\rightarrow\V{i}})$ are random variables sampled from the multinomial distribution with $\sum_{\V{i}'}N_{s,\V{i}\rightarrow\V{i}'}=n_{s,\V{i}}(t)\D{V}$ total trials and probabilities  $(p_s,p_s,\cdots,p_s,p_s,1-2dp_s)$ where $p_s = D_s\D{t}/\D{x}^2$, where we have assumed $\D{x}_1=\cdots=\D{x}_d=\D{x}$.

For fast diffusion, this method becomes more efficient than treating hoping events one by one (as in ISSA).
However, it is an approximate method since the actual distribution of the outward fluxes deviates from the multinomial distribution as $\D{t}$ increases.
In fact, $\D{t}$ cannot be arbitrarily large and is limited by condition~\eqref{stab_cond_beta} because of the requirement $1-2dp_s\ge0$.
We also note that the number of molecules in a cell never becomes negative due to the constraint $\sum_{\V{i}'}N_{s,\V{i}\rightarrow\V{i}'}=n_{s,\V{i}}(t)\D{V}$.
Hence, the fluxes on disjoint faces are correlated, which is different from the FHD description~\eqref{SODE_diff_only}.
In the deterministic limit this scheme converges to a standard forward Euler scheme for the diffusion equation, and is therefore only first-order accurate in time.
In the stochastic setting, this scheme adds correlations between the fluxes through different faces of a given cell in such a way as to ensure discrete fluctuation-dissipation balance for \emph{any} allowable time step size.
In fact, for a system with diffusion only, this method ensures that the equilibrium fluctuations are strictly Poisson, as desired for independent Brownian walkers.

In order to handle chemical reactions, we use the SSA algorithm locally and independently in each cell, without any diffusive events.
Let $\mathfrak{R}_s(\V{n},\tau)$ denote the (random) change in the number density of species $s$ when a cell with initial state $\V{n}$ is simulated using SSA over a time interval $\tau$.
In the absence of diffusion, the SSA-based reaction scheme can be written as
\begin{equation}
\label{SSA}
n_{s,\V{i}}(t+\D{t}) = n_{s,\V{i}}(t)+\mathfrak{R}_s(\V{n}_{\V{i}}(t),\D{t}).
\end{equation}

If we combine the diffusion-only~\eqref{MN} and reaction-only~\eqref{SSA} schemes using a Strang splitting approach~\cite{Strang1968}, we obtain the SSA/2+MN+SSA/2 scheme
\begin{subequations}
\label{SSA2MNSSA2}
\begin{align}
&n_{s,\V{i}}^{\star} = n_{s,\V{i}}^k + \mathfrak{R}_s(\V{n}_{\V{i}}^k,\D{t}/2),\\
&n_{s,\V{i}}^{\star\star} = n_{s,\V{i}}^\star + \mathfrak{D}_{\V{i}}(n_s^\star,\D{t}),\\
&n_{s,\V{i}}^{k+1} = n_{s,\V{i}}^{\star\star} + \mathfrak{R}_s(\V{n}_{\V{i}}^{\star\star},\D{t}/2),
\end{align}
\end{subequations}
where superscripts denote time step or intermediate stage.
(We note that a number of different splitting variants are possible; the version presented here gave the most
accurate structure factor.)
This split scheme has a number of advantages.
It becomes an exact sampler (solver) for the RDME in the limit $\D{t}\rightarrow 0$, just like ISSA.
It is notably more efficient than ISSA if there are many events per time interval $\D{t}$, and it can be parallelized in a straightforward manner using domain decomposition.
Since the number of molecules is always a nonnegative integer both in multinomial diffusion sampling and in SSA, this property is also preserved for this scheme.
Moreover, since both sampling methods preserve the thermodynamic equilibrium distribution (i.e., the Poisson statistics), the split scheme also preserves it for any allowable time step size.

However, this scheme has some disadvantages.
First, the time step size restriction~\eqref{stab_cond_beta} for diffusion becomes severe for fast diffusion.
Second, SSA exhibits poor scalability with respect to the number of molecules in a cell.
This can be resolved by replacing SSA with the tau leaping method, but the nonnegativity is no longer guaranteed.
Third, since the multinomial diffusion method used here is only first-order accurate, the accuracy of the scheme is first order even though Strang splitting is used.
Constructing RDME-based diffusion methods that are more accurate is possible~\cite{LampoudiGillespiePetzold2009,DrawertLawsonPetzoldKhammash2010} but nontrivial and is not the focus of our work.

\subsection{\label{subsec_struct_factor_analysis_RDME}Structure Factor Analysis}

\begin{figure}
\includegraphics[width=0.6\linewidth]{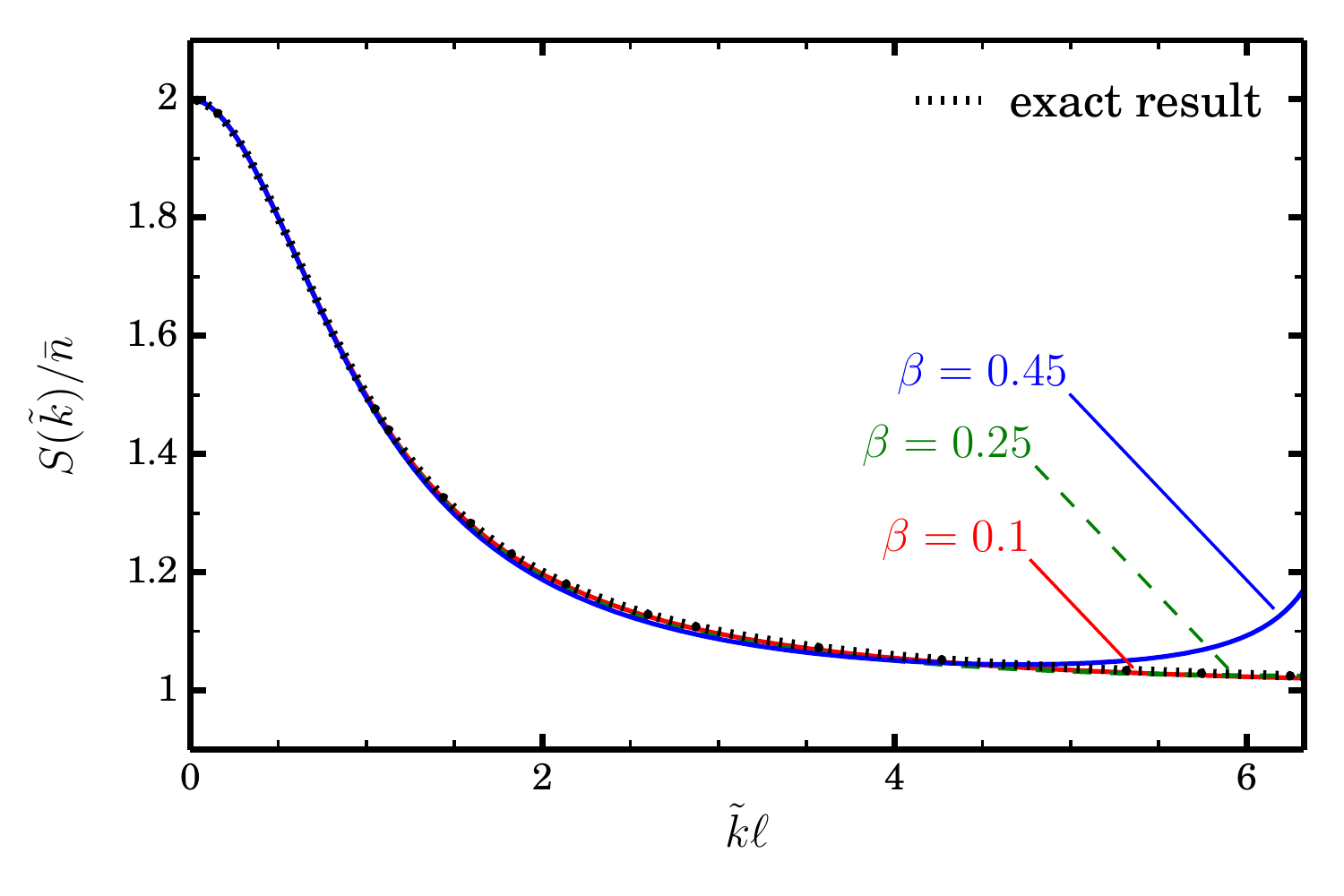}
\caption{\label{fig_Sk_analysis_RDME}
Structure factors $S(\tilde{k})$ obtained from the SSA/2+MN+SSA/2 scheme~\eqref{SSA2MNSSA2} for the one-dimensional linearized FHD equation~\eqref{linearized_SPDE} with $\bar{\Gamma}/\bar{n} r=2$ (e.g., an out-of-equilibrium monostable Schl\"ogl model). Different values of $\beta$ are compared, with $\alpha=0.1\beta$, see Eq.~\eqref{alpha_beta}.
The exact result~\eqref{Sk_exact} is depicted by the dotted line.
Note that the modified wavenumber $\tilde{k}$ is used, see Eq.~\eqref{eq_k_tilde_def}.}
\end{figure}

In this section, we investigate the stochastic accuracy of the SSA/2+MN+SSA/2 scheme~\eqref{SSA2MNSSA2}.
To this end, we consider the structure factor $S(k)$ of an out-of-equilibrium monostable Schl\"ogl model in one dimension (see Sections~\ref{subsec_struct_factor} and \ref{subsec_Schlogl_bg}).
In the limit of many molecules per cell, an asymptotic expression of $S(k)$ can be obtained for the scheme as a function of $\D{x}$ and $\D{t}$.
The multinomial fluxes can be approximated by correlated Gaussian ones and the type of analysis summarized in Appendix~\ref{appendix_linearized_eqn_analysis} can be applied; we do not give the details here for brevity. 
Note that we present a similar structure factor analysis for our FHD-based schemes with some background and details in Section~\ref{subsec_struct_factor_analysis}.

Figure~\ref{fig_Sk_analysis_RDME} illustrates how $S(k)$ deviates from the exact result~\eqref{Sk_exact} as $\D{t}$ is increased to $\D{t}_\mathrm{max}$ (equivalently, to $\beta=0.5$), for $\alpha=0.1\beta$, see Eq.~\eqref{alpha_beta}.
While $S(k)$ is accurately reproduced at the reaction-dominated scales $k\ell\ll 1$ for all values of $\beta$, it becomes inaccurate for smaller scales as $\D{t}$ approaches $\D{t}_\mathrm{max}$.
We recall that for a system at thermodynamic equilibrium, the split scheme \emph{exactly} preserves the correct equilibrium distribution for any $\D{t}<\D{t}_\mathrm{max}$.
This property ensures that, for $\alpha\ll\beta$, good structure factors are obtained even for systems outside of thermodynamic equilibrium, which exhibit a nonzero correlation length. 
For example, for $\beta=0.25$, the SSA/2+MN+SSA/2 scheme in Fig.~\ref{fig_Sk_analysis_RDME} gives a notably more accurate $S(k)$ than the FHD-based \emph{explicit} schemes in Fig.~\ref{fig_Sk_analysis_FHD}.
Hence, even though this split scheme is found to give only first-order accurate $S(k)$, the resulting structure factor is relatively insusceptible to increasing $\D{t}$ until the stability limit is approached.

\section{\label{appendix_averaging_function}Averaging Function $\tilde{n}$}

In this appendix we show that the arithmetic mean should be chosen as the averaging function $\tilde{n}(n_1,n_2)$.
Here we consider the diffusion-only case for a single species and investigate the equilibrium distribution of the spatially discretized FHD equation~\eqref{SODE_diff_only}.
Since the $true$ equilibrium distribution for a bulk (infinite) system is known to be a product Poisson distribution from the corresponding microscopic system consisting of molecules undergoing independent Brownian motions, we choose $\tilde{n}$ so that the resulting equilibrium distribution is as close as possible to the true distribution.
In addition, since the prevention of negative cell number densities is one of the essential issues for the development of a robust FHD numerical scheme, special care is taken to modify the form of $\tilde{n}(n_1,n_2)$ for small values of $n_1$ and $n_2$.
Here we focus on the continuous-time case and do not assume any specific temporal integrator.

The corresponding microscopic system has $N_\mathrm{c}\bar{n}\D{V}$ molecules, where $\bar{n}$ is the mean number density and $N_\mathrm{c}$ is the number of cells.
The equilibrium distribution of the numbers of molecules in each cell follows the multinomial distribution with equal probabilities $1/N_\mathrm{c}$.
Hence, it is straightforward to obtain the following second-order statistics of cell number density $n_{\V{i}}$:
\begin{align}
\label{micro_sys_res_var}
&\Var{n_{\V{{i}}}}=\frac{N_\mathrm{c}-1}{N_\mathrm{c}}\frac{\bar{n}}{\D{V}},\\
\label{micro_sys_res_cov}
&\Cov{n_{\V{{i}}_1}}{n_{\V{{i}}_2}}=-\frac{1}{N_\mathrm{c}}\frac{\bar{n}}{\D{V}}\quad\mbox{for $\V{{i}}_1\ne\V{{i}}_2$}.
\end{align}
Equivalently, from Eqs.~\eqref{micro_sys_res_var} and \eqref{micro_sys_res_cov}, the structure factor is also obtained as
\begin{align}
\label{micro_sys_res_Sk}
S(\V{k})=\bar{n}\quad\mbox{for nonzero $\V{k}$}.
\end{align}

If the FHD system~\eqref{SODE_diff_only} attains an equilibrium state, it can be shown that its second-order statistics are completely characterized by $\av{\tilde{n}}$, which is the equilibrium average of $\tilde{n}$ over all faces:
\begin{align}
\label{analytic_res_var}
&\Var{n_{\V{{i}}}}=\frac{N_\mathrm{c}-1}{N_\mathrm{c}}\frac{\av{\tilde{n}}}{\D{V}},\\
\label{analytic_res_cov}
&\Cov{n_{\V{{i}}_1}}{n_{\V{{i}}_2}}=-\frac{1}{N_\mathrm{c}}\frac{\av{\tilde{n}}}{\D{V}}\quad\mbox{for $\V{{i}}_1\ne\V{{i}}_2$},\\
\label{analytic_res_Sk}
&S(\V{k})=\av{\tilde{n}}\quad\mbox{for nonzero $\V{k}$}.
\end{align}

Comparing Eqs.~\eqref{analytic_res_var}--\eqref{analytic_res_Sk} to the correct result Eqs.~\eqref{micro_sys_res_var}--\eqref{micro_sys_res_Sk} suggests that one needs to choose $\tilde{n}$ so that $\av{\tilde{n}}$ is as close as possible to $\bar{n}$.
It is easy to see that the arithmetic mean (AM) \emph{would} give the right answer: $\av{\tilde{n}^\mathrm{AM}}=\av{\frac12(n_1+n_2)}=\bar{n}$.
On the other hand, to calculate $\av{\tilde{n}}$ for the geometric mean $\tilde{n}^\mathrm{GM}=\sqrt{n_1 n_2}$ or the harmonic mean $\tilde{n}^\mathrm{HM}=2/(n_1^{-1}+n_2^{-1})$, one needs to know the equilibrium distribution $\rho(n_1,n_2)$ of two neighboring cells. 
However, under the reasonable assumption that all three averaging functions give similar distributions $\rho(n_1,n_2)$ allowing only nonnegative number densities, it can be easily shown that $\av{\tilde{n}^\mathrm{HM}}\le\av{\tilde{n}^\mathrm{GM}}\le\av{\tilde{n}^\mathrm{AM}}$ from the well-known inequalities among the Pythagorean means.
In fact, in Fig.~\ref{fig_Schlogl_Sk_rho}, this ordering is observed from the structure factor of the diffusion-only system, see Eq.~\eqref{analytic_res_Sk}. 
Hence, we conclude that the arithmetic mean is the right choice for $\tilde{n}$.

\begin{figure}
\includegraphics[width=0.6\linewidth]{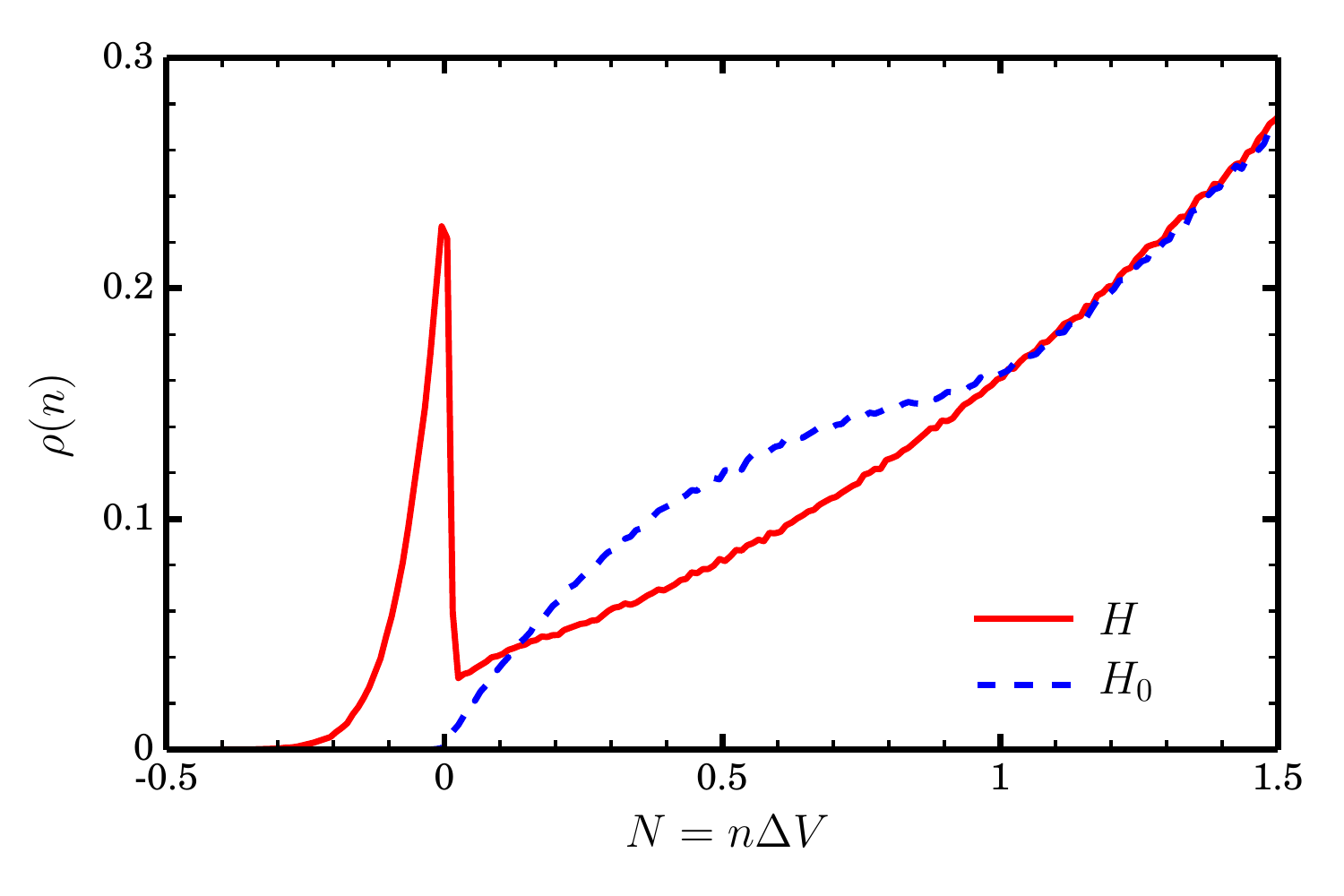}
\caption{\label{fig_hist_near_zero}
The equilibrium cell number density distributions $\rho(n)$ near $n=0$ obtained from the arithmetic mean averaging function~\eqref{avg_type_arithmetic_discont} (depicted by the red solid line), which uses the discontinuous Heaviside function $H$, and Eq.~\eqref{avg_type_arithmetic_smoothed} (depicted by the blue dotted line), which uses the smoothed Heaviside functions $H_0$.
The results are obtained from the one-dimensional diffusion-only FHD system~\eqref{SODE_diff_only} having $D=\bar{n}=\D{x}=1$, $A=5$, and $N_\mathrm{c}=512$ by using the EMTau scheme~\eqref{EMTau} with $\D{t}=10^{-3}$.}
\end{figure}

However, if the arithmetic mean is employed without modification, Eq.~\eqref{SODE_diff_only} does not attain an equilibrium state.
This is because almost surely at some point on some grid face we will have $\tilde{n}<0$ so that the stochastic diffusive flux becomes undefined.
The nonnegativity of cell number densities is guaranteed if the stochastic diffusive flux through a face is turned off when the number density of either cell sharing the face becomes zero.  
Specifically, under some technical assumptions, it can be proven that if $\tilde{n}(n_1,n_2)$ is a nonnegative function that satisfies  
\begin{equation}
\label{cond_nonneg}
\tilde{n}(n_1,n_2)=0\quad\mbox{for $n_1\le0$ or $n_2\le0$},
\end{equation}
then the number density of each cell never becomes negative.
The nonnegative arithmetic mean averaging function
\begin{equation}
\label{avg_type_arithmetic_discont}
\tilde{n}(n_1,n_2)=
\begin{cases}
\frac12 (n_1+n_2) & \mbox{if $n_1>0$ and $n_2>0$,}\\
0 & \mbox{otherwise,}
\end{cases}
\end{equation}
does not allow negative density (i.e., $\rho(n)=0$ for $n<0$).
However, due to the discontinuity of $\tilde{n}$ at $n_1=0$ or $n_2=0$, $\rho(n)$ does not decrease to zero as $n$ becomes zero (i.e., $\lim_{n\rightarrow0^+}\rho(n)>0$) and a delta function is formed
at $n=0$, see Fig.~\ref{fig_hist_near_zero}.

To understand this behavior, we consider numerically integrating
Eq.~\eqref{SODE_diff_only} with a small time step size $\D{t}>0$.
Even if the density in a given cell $n_1(t)$ has a small positive value, $\tilde{n}$ can be large if the density in the neighboring cell $n_2(t)$ is large.
In this case, $n_1(t+\D{t})$ can become negative due to the stochastic diffusive flux.
But, once $n_1$ has become negative, the stochastic diffusive flux is turned off, and due to the deterministic diffusion, $n_1$ increases and becomes positive again. 
Hence, as shown in Fig.~\ref{fig_hist_near_zero}, $\rho(n)$ attains a peak in the negative density region near $n=0$.
The width and height of the peak are proportional to $\D{t}^{1/2}$ and $\D{t}^{-1/2}$, respectively, and the peak becomes a delta function in the limit $\D{t}\rightarrow0$.

We note that the averaging in Eq.~\eqref{avg_type_arithmetic_discont} can be expressed as $\tilde{n} = \frac12(n_1+n_2)H(n_1\D{V})H(n_2\D{V})$ where $H$ is the Heaviside function.
To avoid a discontinuity in $\tilde{n}$ we can use a smoothed Heaviside function to arrive at Eq.~\eqref{avg_type_arithmetic_smoothed}.
Note that the smoothing is based on the number of molecules in a cell ($N=n\D{V}$) and the smoothing region $0\le N\le 1$ is chosen so that the stochastic diffusive flux is modified only when there is less than one molecule in a cell. 
In Fig.~\ref{fig_hist_near_zero}, we show the distribution $\rho(n)$ near $n=0$ obtained by the averaging function~\eqref{avg_type_arithmetic_smoothed}, for a rather small mean number of molecules per cell, $\bar{n}\D{V}=5$.
With the use of a smoothed Heaviside function, the spurious delta function at $n=0$ as $\D{t}\rightarrow0$ is removed, and the probability of negative density is greatly reduced for small $\D{t}$.

\section{\label{appendix_linearized_eqn_analysis}Linearized Equation Analysis}

In this appendix we summarize how the discrete structure factor is obtained as a function of $\D{x}$ and $\D{t}$ when a given spatiotemporal discretization is applied to the linearized FHD equation~\eqref{linearized_SPDE}, following the Fourier-space analysis developed in Ref.~\cite{DonevVandenEijndenGarciaBell2010}.
For simplicity, we consider here the one-dimensional case.

Applying the spatial discretization given in Section~\ref{sec_spatial_discretization} to Eq.~\eqref{linearized_SPDE} and taking a discrete Fourier transform, we obtain an Ornstein--Uhlenbeck equation for the Fourier coefficient $\d{\hat{n}_k}(t)$,
\begin{equation}
\label{linearized_FT_spat_disc}
\ddt\d{\hat{n}_k} = -D\tilde{k}^2\d{\hat{n}_k}+\sqrt{\frac{2D\bar{n}\tilde{k}^2}{V}}\mathcal{W}_k^\paren{\mathrm{D}}-r\d{\hat{n}_k}+\sqrt{\frac{2\bar{\Gamma}}{V}}\mathcal{W}_k^\paren{\mathrm{R}},
\end{equation}
where the modified wavenumber $\tilde{k}$ is defined in Eq.~\eqref{eq_k_tilde_def}, and $\mathcal{W}_k^\paren{\mathrm{D}}(t)$ and $\mathcal{W}_k^\paren{\mathrm{R}}(t)$ are independent standard GWN processes.
Compared to the continuous-space case (see Eq.~\eqref{linearized_FT}), $k$ is replaced by $\tilde{k}$ due to the discrete Laplacian and divergence operators.
Note that in this linearized analysis Poisson processes have been replaced by Gaussian ones with the mean evaluated at the ensemble average (i.e., macroscopic) values of the density.
For convenience, we have replaced complex-valued GWN noise processes by real-valued ones having the same noise intensities.

When the EMTau scheme~\eqref{EMTau} is used to solve Eq.~\eqref{linearized_FT_spat_disc}, we have the recursion
\begin{equation}
\label{dhat_EM}
\d{\hat{n}_k}(t+\D{t})=\left[1-(D\tilde{k}^2+r)\D{t}\right]\d{\hat{n}_k}(t)+\sqrt{\frac{2D\bar{n}\tilde{k}^2\D{t}}{V}}W_1+\sqrt{\frac{2\bar{\Gamma}\D{t}}{V}}W_2,
\end{equation}
where $W_1$ and $W_2$ are independent standard normal random variables.
We write this as
\begin{equation}
\label{dhatnktdt}
\d{\hat{n}_k}(t+\D{t})\:\equald\:M_k\d{\hat{n}_k}(t) + N_k W,
\end{equation}
where $\equald$ denotes being equal in distribution.
For any given temporal integrator we can straightforwardly obtain analytic expressions for $M_k$ and $N_k N_k^*$.
For example, for the EMTau scheme,
\begin{subequations}
\begin{align}
&M_k = 1-(D\tilde{k}^2+r)\D{t},\\
&N_k N_k^* = \frac{2}{V}(D\bar{n}\tilde{k}^2+\bar{\Gamma})\D{t}.
\end{align}
\end{subequations}
The covariance of the noise $N_k N_k^*$ for multinomial diffusion~\eqref{MN} can most easily be obtained from Eq.~\eqref{SkMkNksq} from the observation that, in the absence of reactions, the exact structure factor $S(k)=\bar{n}$ is obtained for any stable time step.

A similar procedure is applicable to numerical schemes having SSA for reactions.
Since SSA is an exact integrator, the linearized reaction part in Eq.~\eqref{linearized_FT_spat_disc} is exactly solved.
For example, for the EM-SSA scheme~\eqref{EMSSA}, we have (cf.\ Eq.~\eqref{dhat_EM})
\begin{equation}
\d{\hat{n}_k}(t+\D{t})=\left[-D\tilde{k}^2\D{t}+e^{-r\D{t}}\right]\d{\hat{n}_k}(t)+\sqrt{\frac{2D\bar{n}\tilde{k}^2\D{t}}{V}}W_1+\sqrt{\frac{\bar{\Gamma}(1-e^{-2r\D{t}})}{r V}}W_2.
\end{equation}
While the expressions of $M_k$ and $N_k N_k^*$ become complicated for the predictor-corrector midpoint schemes, a theoretical analysis is still tractable with the help of symbolic algebra tools.

By calculating $M_k$ and $N_k N_k^*$ for a given numerical scheme, the stability condition and the structure factor can be obtained as follows.
From the condition that the amplification factor $M_k$ should satisfy 
\begin{equation}
\label{amp_fac_cond}
\abs{M_k}\le 1\quad\mbox{for all $k$},
\end{equation}
the stability condition is obtained.
For the EMTau scheme, we obtain
\begin{equation}
\label{stab_cond_EM}
\frac{D\D{t}}{\D{x}^2}+\frac{r\D{t}}{4}\le\frac12.
\end{equation}
The analytic expression for $S(k)=V\av{\d{\hat{n}_k}\d{\hat{n}_k^*}}$ can be calculated from
\begin{equation}
\label{SkMkNksq}
S(k)=\frac{V N_k N_k^*}{1-M_k^2},
\end{equation}
which is obtained from the time invariance relation $\av{\d{\hat{n}_k}(t)\d{\hat{n}^*_k}(t)}=\av{\d{\hat{n}_k}(t+\D{t})\d{\hat{n}^*_k}(t+\D{t})}$ and Eq.~\eqref{dhatnktdt} \cite{DonevVandenEijndenGarciaBell2010}.
From the analytical expressions of $S(k)$ and $S_0(k)=\lim_{\D{t}\rightarrow0}S(k)$, $[S(k_\mathrm{max})-S_0(k_\mathrm{max})]/S_0(k_\mathrm{max})$ is easily obtained, the series expansion of which for small $\alpha$ (and fixed $\beta$) gives Eq.~\eqref{SkSktilde} for the ImMidTau scheme.
Similarly, series expansions for small $\D{t}$ (i.e., fixed ratio $\alpha/\beta$) reveals the temporal order of accuracy of $S(k)$.


\end{document}